\documentclass{aa}  

\usepackage[]{hyperref}
\usepackage{xcolor}
\usepackage{graphicx}
\usepackage{txfonts}
\usepackage{lipsum}
\usepackage{lscape}             
\usepackage{placeins}           
\usepackage{gensymb}

\newcommand{\dn}{$\rm D4000_n$}
\newcommand{\hb}{H$\beta$}
\newcommand{\hdg}{H$\delta_A$+H$\gamma_A$}
\newcommand{\mgfep}{$\rm [MgFe]^\prime$}

\newcommand{\mgtwofe}{$\rm [Mg_2Fe]$}
\newcommand{\mgfef}{$\rm [MgFe50]^\prime$}
\newcommand{\mgfeft}{$\rm [MgFe52]^\prime$}
\newcommand{\micron}{$\mu m$}

\newcommand{\silver}{{\it silver}}
\newcommand{\golden}{{\it golden}}

\begin{document}
\title{LEGA-C stellar populations scaling relations.}
\subtitle{II - Dissecting mass-complete archaeological trends and their evolution since $z\sim0.7$~with LEGA-C and SDSS}

\author{Anna R. Gallazzi\inst{1}\thanks{anna.gallazzi@inaf.it}
\and Stefano Zibetti\inst{1}
\and Arjen van der Wel\inst{2}
\and Angelos Nersesian\inst{2,3}
\and Yasha Kaushal\inst{4}
\and Rachel Bezanson\inst{4}
\and Daniele Mattolini\inst{1,5}
\and Eric F. Bell\inst{6}
\and Laura Scholz-Diaz\inst{1}
\and Joel Leja\inst{7,8,9}
\and Francesco D'Eugenio\inst{10,11}
\and Po-Feng Wu\inst{12,13,14}
\and Camilla Pacifici\inst{15}
\and Michael Maseda\inst{16}}
\institute{INAF-Osservatorio Astrofisico di Arcetri, Largo Enrico Fermi 5, 50126, Firenze, Italy
\and Sterrenkundig Observatorium Universiteit Gent, Krijgslaan 281 S9, B-9000 Gent, Belgium
\and STAR Institute, Université de Liège, Quartier Agora, Allée du six Aout 19c, B-4000 Liege, Belgium
\and Department of Physics and Astronomy and PITT PACC, University of Pittsburgh, Pittsburgh, PA 15260, USA
\and Dipartimento di Fisica, Universit\`a di Trento, Via Sommarive 14, I-38123 Povo (TN), Italy
\and Department of Astronomy, University of Michigan, 1085 South University Avenue, Ann Arbor, MI 48109, USA
\and Department of Astronomy and Astrophysics, 525 Davey Lab, The Pennsylvania State University, University Park, PA 16802, USA
\and Institute for Gravitation and the Cosmos, The Pennsylvania State University, University Park, PA 16802, USA
\and Institute for Computational and Data Sciences, The Pennsylvania State University, University Park, PA 16802, USA
\and Kavli Institute for Cosmology, University of Cambridge, Madingley Road, Cambridge CB3 0HA, UK 
\and Cavendish Laboratory – Astrophysics Group, University of Cambridge, 19 JJ Thomson Avenue, Cambridge CB3 0HE, UK 
\and Graduate Institute of Astrophysics, National Taiwan University, Taipei, 10617, Taiwan
\and Department of Physics and Center for Theoretical Physics, National Taiwan University, Taipei, 10617, Taiwan
\and Physics Division, National Center for Theoretical Sciences, Taipei, 10617, Taiwan
\and Space Telescope Science Institute, 3700 San Martin Drive, Baltimore, MD 21218, USA
\and Department of Astronomy, University of Wisconsin-Madison, 475 N. Charter St., Madison, WI, 53706, USA}

\abstract{

We analyse a sample of 552  galaxies from the LEGA-C spectroscopic survey ($0.6 < z < 0.77$), for which we estimated stellar population parameters by a Bayesian analysis of stellar absorption features and photometry. We investigate how current star formation activity influences light-weighted mean stellar ages and metallicities, and their median trends with stellar mass or velocity dispersion. The bimodality in the global age–mass relation stems from the different age distributions in the quiescent and star-forming populations. A bimodality is not observed in the stellar metallicity-mass relation, although quiescent and star-forming galaxies have different distributions in this parameter space. We identify a high-metallicity sequence populated by both quiescent and weakly star-forming galaxies. At masses below $10^{10.8}M_\odot$ the median stellar metallicity-mass relation of star-forming galaxies steepens,  as a consequence of increasing scatter toward lower stellar metallicities for galaxies with increasing specific star formation rate at fixed mass. 
Relying on a consistent analysis of SDSS DR7 spectra, accounting for aperture corrections, we quantify the evolution of the volume-weighted stellar age and stellar metallicity scaling relations between z=0.7 and the present. We find negligible evolution in the stellar metallicity-mass relation of quiescent galaxies and for $M_\ast > 10^{11} M_\odot$ galaxies in general. Lower mass star-forming galaxies, instead, have typically lower metallicities than their local counterparts, indicating significant enrichment since $z\sim0.7$ in the low-mass regime. Notably, the median of the stellar ages of both the general population and quiescent galaxies has changed by only 2~Gyr between z=0.7 and z=0.1, less than expected from cosmic aging. Some quiescent galaxies must evolve passively to reach the old boundary of the local population. However, in order to explain the evolution of the median trends, both individual evolution, through rejuvenation and/or minor merging impacting the outer galaxy regions, and population evolution, through quenching of massive, metal-rich star-forming galaxies, are required.}

   \maketitle

\section{Introduction} \label{sec:intro}
The evolution of the cosmic star formation rate density indicates a pervasive decline of star formation since $z=1-2$ \citep{MD14,Driver18}. The evolution of the mass and number densities of passive and active galaxies indicate that this happens together with the build-up of the massive quiescent population \citep{Muzzin13b,Ilbert13,Tomczak14,leja20}. Nevertheless, massive evolved systems have been found and spectroscopically confirmed at $z=2-4$ \citep{Cimatti04,Daddi05,Fontana09,Toft12,Whitaker13,Glazebrook17,Valentino20,DEugenio21} and are now continuously discovered at increasing redshift ($z>4$ up to $\sim7$) with the sensitive near-IR capabilities of JWST \citep{Valentino23,Carnall23,Carnall24,DEugenio24,Glazebrook24,Weibel25,deGraaff25}. Models of galaxy formation and evolution need to reproduce both the number densities of passive galaxies at increasing redshift and the global star formation suppression in more recent epochs.

Complementary and crucial information about the efficiency of star formation and the quenching mechanisms can come from studying the physical properties of the stellar populations in galaxies, in addition to the gas component. The metal content in galaxies is the result of metal production during star formation, gas and metal loss through SN- or AGN-driven outflows, and gas infall from the circum-galactic medium (CGM) and inter-galactic medium (IGM), as well as matter accretion through mergers. Thus, tracing the metallicities of galaxies with different global properties, in combination to recent and past star formation histories, can provide constraints to the mechanisms regulating star formation efficiency and quenching \citep[see for a review][]{MM19}.

In particular, the metallicity of the stellar populations in galaxies, averaged over the galaxy star formation history (SFH), is representative of the chemical enrichment at the peak epoch of mass build-up, which is to first order the (light- or mass-weighted) average age of the stellar populations. Both of these fossil record properties are known to scale with galaxy mass \citep[e.g.][]{kauffmann03,gallazzi05,mateus06,panter08,Trussler21}. Not only are they indicative of the \textit{past} star formation and assembly histories, but also they have been shown to depend on the \textit{present-day} star formation activity of galaxies. The chemistry of galaxies may thus independently suggest a correlation between the present-day star formation activity and the long-timescale star formation histories \citep[e.g.][]{Chauke18,Mattheet18}.

In the local Universe, the spectral quality and statistics of surveys such as SDSS have made it possible to show that quiescent and star-forming galaxies follow different relations between their ages, stellar metallicities \citep[][]{gallazzi05,peng15,Trussler21,Gallazzi21,Mattolini25}, as well as $\alpha$/Fe, with galaxy mass \citep{Gallazzi21}. It is now established that quiescent galaxies follow tight relations of increasing age, metallicity and $\alpha$/Fe with mass \citep[e.g.][]{worthey92,trager2000,bernardi05,Thomas05,gallazzi06,graves09a,Renzini06,mcdermid15}. While differing in details, there is general agreement in finding lower stellar metallicities in star-forming galaxies with respect to quiescent galaxies of similar mass, more significantly for lower-mass galaxies. At face value, these trends suggest a more prolonged and less efficient metal enrichment in galaxies with on-going star formation. 

However, what is the physical origin is still a matter of debate. Empirical works and comparison with analytical models support a scenario in which starvation, i.e. suppression of fresh gas supply, is the dominant quenching route in massive galaxies \citep{peng15,Trussler20}. However, outflows are also suggested to be important contributors, or even the only mechanism, driving the lower stellar metallicities of low-mass star-forming galaxies \citep{spitoni17}. The reduced effective yield observed in lower mass galaxies indeed seem to require SN-driven metal-rich outflows, coupled with large gas fractions and low star formation efficiency \citep{Dalcanton07}. The different global properties of quiescent and star-forming galaxies may also originate from a universal relation with the local conditions, namely the local stellar mass density, which drives both the efficiency of star formation and the ability of a galaxy to accumulate and retain metals \citep{BdJ00,kauffmann03b,Zhuang18,Neumann21,Zibetti22,Vaughan22}, and the different structural properties of galaxies. 

More in general, a secondary dependence of the metallicity of galaxies with the current star formation rate is observed not only for the gas-phase component, as established by several works \citep[e.g.][]{Mannucci10,LaraLopez10,Hunt12}, but also for the stellar component \citep{Looser24}. Such a dependence is also a prediction of both semi-analytic models \citep{fontanot21} and cosmological hydrodynamical simulations \citep{Mattheet18,Garcia24} and it is associated to the efficiency of stellar feedback and a regulation of the timescales for  inter-stellar medium (ISM)  enrichment and star formation. Thus, connecting the metallicity of the stars with the star formation activity and the chemical enrichment of the ISM can constrain the gas and metal flow in and outside galaxies \citep{Dalcanton07,Lu15,Lian18}.

The scaling relations observed in the local Universe give us a picture of the present-day population, but comparing galaxies of similar mass does not mean comparing descendants with their progenitors. Crucial information to disentangle individual evolutionary paths from population evolution comes from tracing these fossil record scaling relations at increasing redshift. This would eventually allow a statistical connection between progenitors and descendants, and increase our ability to resolve the early phases of galaxy formation. 
Various works tracing the metal abundances in stars for quiescent galaxies at intermediate redshift, both in the field and in clusters \citep{Schiavon06,ferreras09,SB09,choi14,gallazzi14,jorgensen17} generally agree on mild or absent metallicity evolution of the quiescent population. These results have been corroborated with larger samples from deep spectroscopic surveys such as LEGA-C \citep{Beverage23,Bevacqua24} and VANDELS \citep{saracco23}, as well as for the most massive galaxies at higher redshifts \citep{Kriek19}.
Contrary to the apparent lack of metallicity evolution, evidence for non-passive evolution of the population of quiescent galaxies comes from their ages (or mass-weighted formation epochs), inferred from a variety of sources and methods: from stellar population analyses of high-resolution spectra and spectral absorption features \citep{Schiavon06,SB09,gallazzi14,kaushal24} or multi-narrow-band photometry and grism spectroscopy \citep{diaz-garcia19,EC19}, or indirectly from evolution of the color-magnitude plane \citep{ruhland09} and the mass-size plane \citep{Belli15}, in comparison to those inferred in the present-day Universe.

Despite that these observational results seem to converge toward a qualitatively similar picture, systematic differences associated to different spectral inference methods, observational diagnostics and sample selection lead to quantitative differences in the detailed shape of the scaling relations and the mass scale at which deviations from individual or population passive evolution are important. This may in turn affect the comparison with model predictions.
To overcome these limitations, it is thus crucial to adopt consistent diagnostics and modeling approaches when comparing galaxies at different redshifts \citep[e.g.][]{gallazzi14,saracco23}. Moreover, although diagnostics of the physical properties of stellar populations are more easily measured and interpreted in quiescent galaxies, it is paramount to extend this type of studies to star-forming galaxies at different epochs as well, in order to include the potential progenitors of present-day galaxies \citep{gallazzi14,Trussler21,kaushal24,Nersesian25}.

In this work, we tackle both issues by relying on a consistent analysis of LEGA-C and SDSS spectra for samples including both quiescent and star-forming galaxies. Our intermediate-redshift sample is drawn from the LEGA-C survey \citep{vdWel16,DR3} to have robust absorption index diagnostics. The resulting sample of $\sim550$ galaxies in the redshift range $0.6<z<0.77$ is representative of the general galaxy population down to $10^{10.4}M_\odot$ and of the star-forming population down to $10^{10}M_\odot$.
We focus on carefully selected spectral absorption features that are chiefly sensitive to age and total stellar metallicity, while being minimally affected by abundance ratio variations and dust. This differs from other works performing full-spectral fitting \citep[e.g.][]{Barone22,Cappellari23,kaushal24,Nersesian25}. These features, coupled with photometry, are interpreted accounting for complex star formation and chemical enrichment  histories and dust attenuations, to derive average ages, stellar metallicities, as well as stellar masses for individual galaxies. Our methodology and resulting parameter estimates are described in depth in \cite{Gallazzi25a} (hereafter Paper I).  

With this dataset we establish the scaling relations of stellar population properties at intermediate redshift for quiescent and star-forming galaxies. We wish to address whether galaxies differ in the physical parameter space depending on their star formation activity at the epoch of observations. Comparing galaxies as a function of stellar mass or as a function of stellar velocity dispersion can give us clues on whether quenching and metal enrichment are mainly regulated by the integral of the star formation history or by the central concentration of total mass (including non-baryonic components).
We combine the LEGA-C analysis with consistent physical parameter estimates for a volume- and completeness-weighted sample of galaxies from the SDSS DR7 \citep{Mattolini25}. Importantly, these SDSS estimates derive from aperture-corrected absorption index strengths for individual galaxies \citep{Zibetti25}. We are thus in the right position to perform a coherent comparison between the stellar population scaling relations at different epochs, $\left<z\right>=0.1$~and $\left<z\right>=0.7$, controlling for potential biases, and quantify the evolution in ages and stellar metallicities for quiescent and star-forming galaxies separately.

The paper is organized as follows. In Sec.~\ref{sec:data} we summarize the sample drawn from the LEGA-C survey and the spectral inference of stellar population parameters as derived in Paper I. Our main results are presented in Sec.~\ref{sec:scaling_sfr}, where we analyse the volume-weighted age and stellar metallicity scaling relations for quiescent and star-forming galaxies, and their possible secondary dependence on star formation rate. The evolution of the scaling relations between $\left<z\right>=0.7$~and $\left<z\right>=0.1$, globally and for quiescent and star-forming galaxies separately, are presented in Sec.~\ref{Sec:evolution} with a coherent LEGA-C and SDSS analysis. We discuss the implications of our results in Sec.~\ref{sec:discussion} and summarize our findings in Sec.~\ref{sec:summary}. 
Throughout the paper we assume a $\Lambda$CDM cosmology with $H_0=70~ \rm km/s/Mpc$, $\Omega_M=0.3$, $\Omega_\Lambda=0.7$, and a solar stellar metallicity of $Z_\odot=0.0154$ and we adopt a \cite{chabrier03} IMF.

\section{The LEGA-C dataset}\label{sec:data}

\subsection{Samples definition}\label{sec:data_sample}
\begin{figure}
    \centering
    \includegraphics[width=1\linewidth]{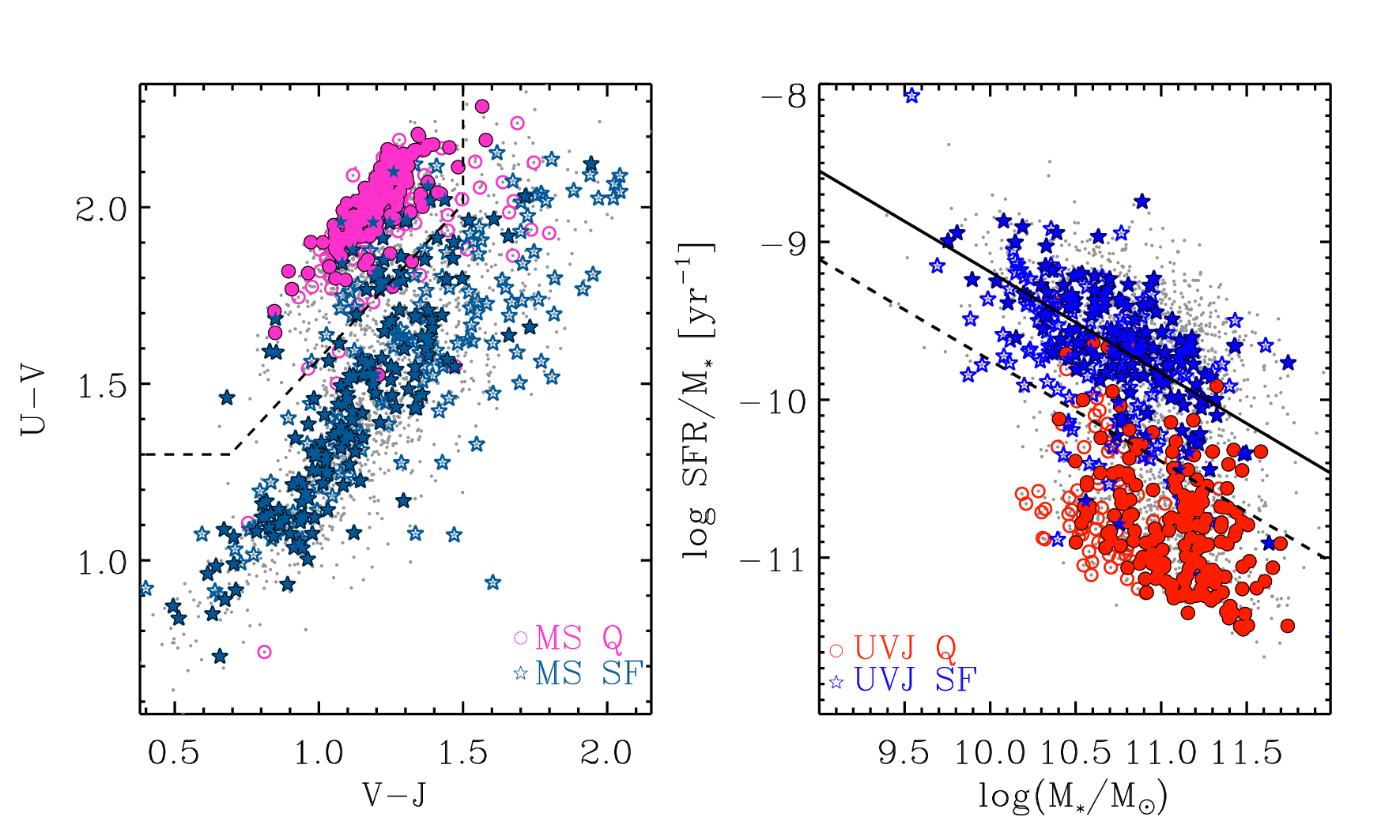}
    \caption{Distribution in U-V versus V-J (left panel) and SSFR versus $M_\ast$ (right panel) for the parent LEGA-C sample (small dots) and for the \silver~(empty symbols) and high-S/N \golden~samples (filled symbols) used in this work. In the left panel, galaxies are distinguished into Q (magenta) and SF (blue) based on their distance from the star-forming main sequence, our default classification (i.e. Q galaxies lie below the dashed line in the right panel). In the right panel, red/blue symbols identify Q/SF galaxies according to their location in the UVJ plane (left and right of the dashed line in the right-hand panel, respectively).}
    \label{fig:UVJ_SSFR_sample}
\end{figure}

Our sample is drawn from the Data Release 3 of the LEGA-C spectroscopic survey \citep{vdWel16,DR3}. We require galaxies to belong to the primary sample, the spectrum to satisfy the criterion $\tt flag\_spec=0$ and $\tt sigma\_star > 0$ and  reliable measurements of selected sets of absorption features for stellar metallicity and age estimates to be available, as described in Paper I. Our final working sample (which we call \silver~sample) comprises 552  unique galaxies, of which 323  have spectral $S/N>20$ and constitute the \golden~sample. The \silver~sample covers the redshift range $0.55<z<0.77$ (as a consequence of wavelength coverage requirement), the stellar mass range $10\lesssim\log(M_\ast/M_\odot)\lesssim 11.5$, and a range in star-formation activity representative of the parent LEGA-C sample. Figure~\ref{fig:UVJ_SSFR_sample} ~illustrates the distribution in the colors $(U-V)-(V-J)$ (the UVJ diagram) and specific star formation rate (SSFR) versus stellar mass ($\rm SSFR-M_\ast$) of the \silver~sample and the \golden~subsample (filled symbols), compared to the parent LEGA-C sample (small dots). As can be seen from the figure, the requirement of high S/N for the \golden~sample tends to disfavor lower-mass galaxies and/or red star-forming galaxies.

We adopt a classification of quiescent (Q) and star-forming (SF) galaxies based on their SSFR. We use  star formation rate (SFR)  estimates based on the UV and 24\micron~fluxes ($\rm SFR_{UVIR}$), following \cite{Whitaker14}, which builds on converting 24\micron~fluxes into total IR luminosity with \cite{DaleHelou} templates, and combining it with UV luminosity as prescribed by \cite{bell05}. These estimates thus account for both unobscured and obscured star formation. The SSFR is obtained dividing the $\rm SFR_{UVIR}$ by our estimate of stellar mass. We fit\footnote{Throughout the paper, to fit parametric relations we use the {\tt MPFIT} IDL routine which finds the bestfit parameters of the supplied function by minimizing the sum of the weighted squared differences between the model and data.} a linear relation between the SSFR and stellar mass for galaxies that are classified as star-forming based on the $UVJ$ diagram \citep{Muzzin13b} (to the right of the dashed line on the left panel of Fig.~\ref{fig:UVJ_SSFR_sample}). The best-fit linear relation is:
\begin{equation}
\log (\rm SFR/M_\ast [yr^{-1}]) = A \cdot (\log (M_\ast/10^{11}M_\odot) + B \label{eqn:ssfr_mass}
\end{equation}
where $A=-0.64\pm0.03$, $B=-9.83\pm0.01 \rm[yr^{-1}]$. To avoid being biased by underestimated SFR, hence erroneously including quiescent galaxies in the scatter of the SF sequence, we compute the one-sided scatter considering only positive deviations, which amounts to $\sigma=0.28$ (the total scatter would be 0.32). We consider as quiescent those galaxies deviating by more than $2\sigma$ below Eq.\ref{eqn:ssfr_mass} (dashed line on the right panel of Fig.~\ref{fig:UVJ_SSFR_sample}). This is not meant to provide a new calibration of the star-forming Main Sequence, which has been derived with larger and more complete datasets \citep[e.g.][]{Whitaker17,Leja22}. Instead, this serves as an operational definition tailored to our sample.

Figure~\ref{fig:UVJ_SSFR_sample} also shows that a cut in SSFR is more conservative for the quiescent population, picking up the reddest galaxies in the UVJ quiescent zone. Conversely, the UVJ passive selection extends to galaxies with higher SSFR. Our \silver~sample comprises 232 Q and 320 SF galaxies. We comment further on different classification criteria in Appendix~\ref{appendix:sys}

\subsection{Physical parameter estimates}\label{sec:data_parameters}
Estimates of stellar mass, r-band-weighted mean stellar metallicity and age are obtained with our {\tt BaStA} code \citep[][Bayesian Stellar population Analysis]{Zibetti17,Zibetti22}. The code performs a Bayesian analysis of key stellar absorption features and photometry with a comprehensive library of simulated spectra constructed by convolving SSP models with randomly generated SFHs, metallicity histories and dust attenuation parameters. The models are based on the Simple Stellar Populations from the 2019 version of the \citet{bc03} (BC03) population synthesis code (CB19), assuming a \cite{chabrier03} IMF, convolved with: i) delayed Gaussian SFHs with stochastic bursts \citep{Sandage1986}, ii) chemical enrichment histories $Z_*(M(t))$ following a generalized leaky box model \citep{Erb08}, and attenuated by dust following \citet{CF00}. A full description of the model library can be found in \cite{Mattolini25} and Paper I. 

As observational constraints, we consider subsets of indices that include at least one age-sensitive index (among \dn, Balmer absorption lines \hb~ and \hdg) and at least one metal-sensitive index (among \mgfep, \mgtwofe, \mgfef, \mgfeft), that have negligible dependence on $\alpha$/Fe. Using galaxies for which all the indices are available, we checked that the chosen subsets of indices predict physical parameter estimates that agree with each other within a few hundredths of a dex, with a scatter comparable to or smaller than the mean parameter uncertainties. Thus we are confident that parameters derived with different indices, among our choices, are consistent within the uncertainties. We use the revised measurements of absorption indices provided in Paper I from the emission-line-subtracted DR3 spectra and accounting for duplicate observations. In addition to absorption indices, we add constraints from the $rizYJ$~photometry from \cite{Muzzin13a}. By comparing these observables with the whole model library, we derive probability density functions of stellar mass, light (and mass)-weighted mean ages and stellar metallicities. We refer the reader to Paper I for details on the methods, the observational constraints and the modeling assumptions, as well as a discussion of potential sources of bias, but we provide here a summary of the bulk uncertainties. 

Statistical uncertainties on individual estimates are on average 0.16~dex for light-weighted age, 0.12~dex for stellar mass, 0.17/0.34~dex for stellar metallicity for \silver~quiescent and star-forming galaxies respectively. We also quantify systematic uncertainties associated with different modeling assumptions within our method, in particular regarding the functional form of the continuous SFH, the metallicity history and the base population synthesis models. We found that stellar metallicity is subject to larger systematics (0.28/0.34~dex for quiescent/star-forming) as a result of changes in SPS models (from BC03 to CB19), chemical enrichment histories (from constant metallicity along the SFH to increasing with the mass formed)  and star formation histories  (from declining exponential to delayed Gaussian), while light-weighted ages can change by $-0.18/-0.13$~dex due to SFH assumptions (being younger assuming a delayed Gaussian functional form).  Moreover, we compare our physical parameter estimates with those obtained on the same sample from indepenent fits with {\tt Prospector} \citep{Nersesian25} and {\tt Bagpipes} \citep{kaushal24}, which adopt different SPS models and assumption on SFH and metallicity.  The offset between different light-weighted age estimates is typically within 0.1 dex, but non constant with age as a consequence of different SFH assumptions. The scatter in age estimates between the three codes is $\sim0.16/0.25$~dex for Q/SF galaxies. The typical uncertainties on light-weighted ages are $0.15/0.15$~dex from {\tt BaStA}, $0.02/0.03$~dex from {\tt Prospector}, $0.01/0.03$~dex from {\tt Bagpipes}. For stellar metallicity the scatter between the estimates from the three codes is $0.24/0.4$~dex for Q/SF galaxies, to be compared with the average uncertainty of $0.16/0.25$~dex from {\tt BaStA}, $0.03/0.04$~dex from {\tt Prospector}, $0.02/0.05$~dex from {\tt Bagpipes}. The uncertainties quantified with {\tt BaStA} thus typically account for more than 60\% of the scatter, compared to $\sim10$\% from {\tt Prospector} or {\tt Bagpipes}. We thus consider uncertainties from {\tt BaStA} to be more representative of the underlying degeneracies. In previous works, we have checked that {\tt BaStA} uncertainties are well-calibrated and consistent with the scatter between input and retrieved parameters in mock spectra \citep{Rossi25}. The reader is referred to Appendix D and E of Paper I for a more in-depth discussion. We comment in Appendix~\ref{appendix:sys} on the robustness in the general trends for quiescent and star-forming galaxies, despite the significant one-to-one variability. The substantial differences in the estimates from different approaches highlight the necessity to adopt the same modeling framework when comparing different redshifts, as we aim to do in this work with a consistent SDSS analysis.

\subsection{Statistical weights}

Finally, we apply statistical weights to correct for incompleteness and selection biases. In particular we consider i) a volume correction, because of the K-band magnitude selection of LEGA-C \citep[the {\tt Vcor} parameter in DR3 from][]{DR3}, ii) a completeness correction to account for the representativeness of the observed LEGA-C sample with respect to the parent sample \citep[the {\tt Scor} parameter in DR3 from][]{DR3}, and finally iii) a spectroscopic correction which accounts for incompleteness in the plane of rest-frame $U-V$ vs absolute g-band luminosity (as observational proxy of physical parameters plane) due to 
our sample selection based on the availability of selected absorption features (see Paper I for more details). Throughout the paper we present as reference scaling relations those obtained by applying these weights to the \silver~sample. We present also the median trends obtained from the un-weighted \golden~sample as a control on higher-quality measurements.

\section{The dependence on the current star formation activity}\label{sec:scaling_sfr}
In Paper I, we have presented the global scaling relations of light-weighted mean stellar age and metallicity. The distribution in age revealed the presence of two sequences transitioning around a stellar mass of $10^{11}M_\odot$ or a velocity dispersion of $\sim230~km/s$. As opposed to age, stellar metallicity shows a shallow increase with stellar mass or velocity dispersion at the massive end, and a steepening at lower masses associated with an increased scatter toward low metallicities, without a clear bimodality.
Here we investigate how the current star formation activity and, in particular, whether a galaxy is quiescent or not and its SSFR, influences the location of $\left<z\right>=0.7$ galaxies in the stellar metallicity-mass and age-mass planes. 
In the local Universe a few studies have shown that quiescent and star-forming galaxies differ not only in their mean ages, as expected from the more prolonged and continuing star formation activity, but also in their stellar metallicities with quiescent galaxies being systematically more metal-rich than equally massive star-forming galaxies below a certain stellar mass \citep[][]{gallazzi05,peng10,Gallazzi21,Trussler20,Mattolini25}. 

Here we explore whether and to what extent the difference in stellar populations of quiescent and star-forming galaxies holds at intermediate redshift. 
We compare scaling relations with stellar mass and velocity dispersion for Q and SF galaxies. We comment on the robustness of our results against different stellar population parameter estimates and  different SFR calibrations in Appendix~\ref{appendix:sys}.

\subsection{Trends with stellar mass}\label{sec:scaling_mass}

\begin{figure*}[!ht]
\centering
\includegraphics[width=0.8\textwidth]{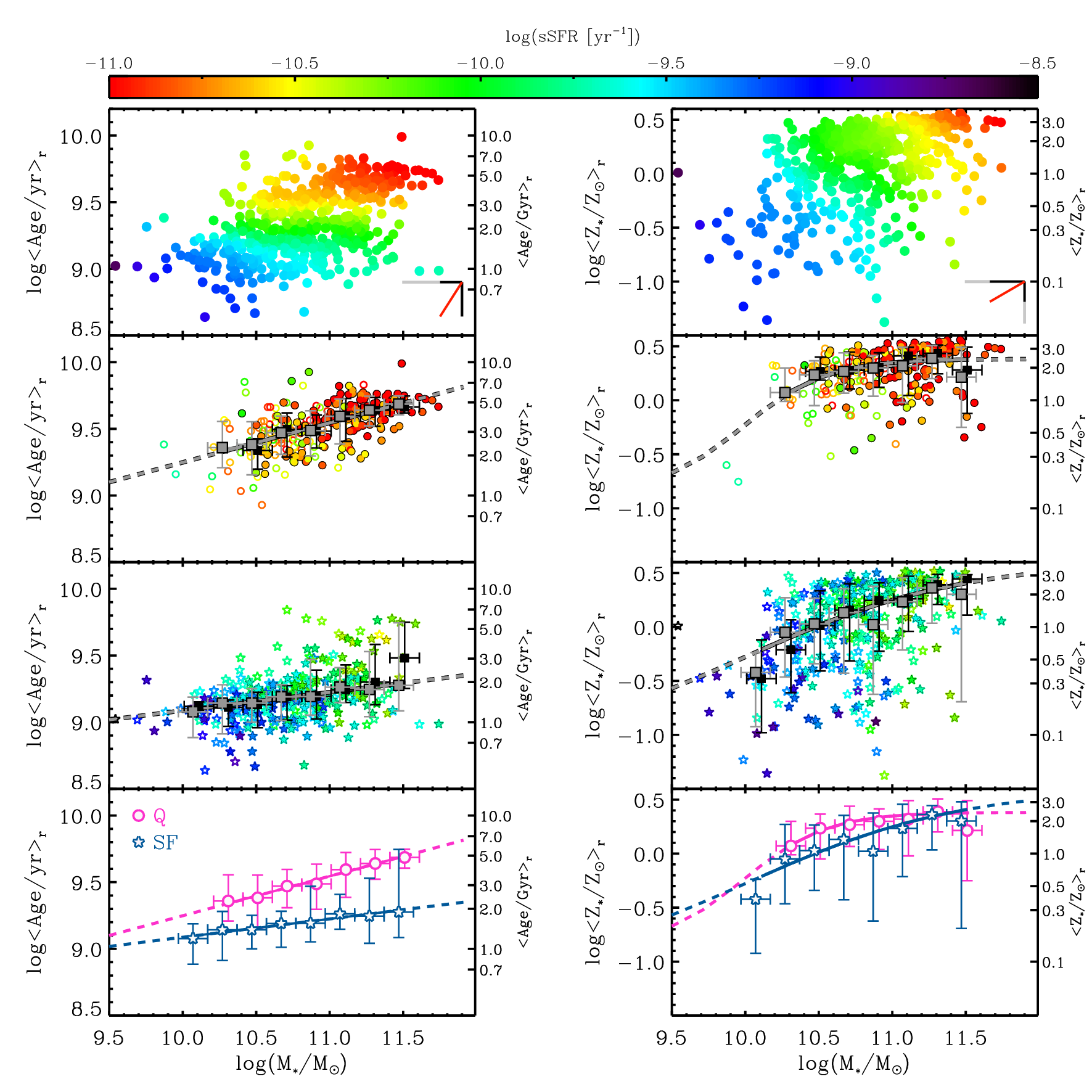}
\caption{Luminosity-weighted mean age (left panels) and mean stellar metallicity (right panels) as a function of stellar mass and of SSFR for LEGA-C galaxies at $z\approx 0.7$. Upper panels: the data points are color coded by SSFR after applying LOESS-smoothing. Black vectors in the lower right corner are proportional to the partial correlation coefficients computed as in \cite{ScholzDiaz24}, and the red vector indicates the direction of maximal increase of SSFR. The grey vectors correspond to a reference correlation coefficient of 0.7. Middle panels: Quiescent galaxies (circles) and star-forming galaxies (stars), color coded by their SSFR, without smoothing. Filled symbols highlight galaxies in the \golden~subsample. Filled squares show the median age and median metallicity in bins of stellar mass 0.2~dex wide and with at least 5 galaxies (grey for the weighted \silver~sample, black for the non-weighted \golden~galaxies). The errorbars indicate the 16-84 percentile range of each distribution (errors on the median are smaller).  The median age-mass trends are fit with a linear relation (Eqn.\ref{eqn:age_mass}), while the median  metallicity-mass trends are fit with the functional form as in Eqn.~\ref{eqn:met_mass}. Grey solid lines show the fit to the \silver~median relations, with extrapolation shown by the dashed lines. Bottom panels: direct comparison of the volume and completeness-weighted median relations for quiescent (magenta) and star-forming (blue) galaxies in the \silver~sample.}
\label{fig:scaling_Q_SF_legac}
\end{figure*}
The upper panels of Fig.~\ref{fig:scaling_Q_SF_legac} display the age--mass and stellar metallicity--mass relations for the \silver~LEGA-C sample color-coded for the SSFR, after applying {\tt LOESS}-smoothing\footnote{We use the {\tt CAP\_LOESS\_2D} routine of Cappellari et al. (2013b), which implements the multivariate {\tt LOESS} algorithm of Cleveland\&Devlin (1988), assuming a degree 1 approximation with frac=0.3.}. To better quantify and visualize the variation of SSFR as a function of the stellar population parameters, we computed partial correlation coefficients of the original (non-smoothed) SSFR (black vectors in the lower right corner), following \cite{ScholzDiaz24}. The red vector indicates the direction of maximal increase of SSFR in the age-mass and metallicity-mass planes. We see the expected trend of decreasing SSFR with increasing age at fixed stellar mass. The partial correlation coefficients indicate a stronger correlation of SSFR with age at fixed mass ($\rho=-0.70\pm0.04$) than with stellar mass at fixed age ($\rho=-0.26\pm0.07$). We also detect a gradient of decreasing SSFR along the stellar metallicity--mass relation, though the correlation with metallicity at fixed mass ($\rho=-0.33\pm0.07$) is weaker than with mass at fixed metallicity ($\rho=-0.41\pm0.06$).
The larger statistics of LEGA-C confirm the trends with SSFR that were seen in the smaller sample of \cite{gallazzi14} with different modeling assumptions. 

In the middle panels of Fig.~\ref{fig:scaling_Q_SF_legac}, we show the distributions for Q (circles) and SF (stars) galaxies separately (\golden~galaxies highlighted by filled symbols), with color-coding reflecting the SSFR of individual galaxies. By comparing these with the top panels, we notice that the average trends in SSFR result from the superposition of both quiescent and star-forming galaxies in the high-metallicity, old-age part of the galaxy distribution, and a lack of quiescent galaxies at lower metallicities/younger ages. 

Median ages/metallicities in bins of stellar mass are shown by filled squares (grey for \silver~sample applying volume and completeness weights, black for \golden~sample without weights; see Table~\ref{Tab:mean_relations}). Quiescent and star-forming galaxies follow two different sequences in luminosity-weighted mean age and stellar mass (left panels). Simple linear fits in the form:
\begin{equation}
\log \left<Age_r/yr\right> = \alpha \cdot (\log (M_\ast/10^{11.5}M_\odot) + P_0 \label{eqn:age_mass}
\end{equation}
describe the trends well, indicating a steeper relation for Q with respect to SF galaxies (Table~\ref{Tab:functional_fit_age_Q_SF}). Over about an order of magnitude in stellar mass, the luminosity-weighted mean age of Q galaxies increases by $\sim3$~Gyr, while that of SF galaxies by only $\sim800$~Myr. The Q and SF age sequences remain distinct almost across the whole mass range probed, and they reach similar ages between 1.5 and 2~Gyr at masses below $10^{10.5}M_\odot$  with less than $1\pm0.3$~Gyr difference.

We checked that mass-weighted mean ages show similar trends.  In particular we find that, for both Q and SF galaxies, the slope of their respective age-mass relations are consistent within less than $1\sigma$~when weighing in mass or in light. The zero-point of the relation for Q galaxies is higher by $0.04\pm0.02$~dex when weighing in mass, while for SF galaxies it is higher by $0.15\pm0.04$~dex. These shifts bring the median mass-weighted age of SF galaxies to overlap with that of Q galaxies at $\log(M_\ast/M_\odot)\leq10^{10.7}$.  This suggests that, despite their different (recent) SFHs, the bulk of the stellar populations of Q and SF galaxies below $10^{10.7}M_\odot$ formed at similar epochs. This is not the case for the more massive galaxies for which the mean formation epoch, approximated by the mass-weighted mean age, differs by 3 Gyr  ($\Delta(\log\left<Age\right>)=0.41\pm0.14$ ~dex)  between Q and SF galaxies.

Quiescent galaxies follow a very shallow mass--metallicity relation, with an increase in median metallicity of 0.15 dex over an order of magnitude in stellar mass. We notice an increased scatter toward low metallicities  in the most massive bin. This is associated with few massive Q galaxies whose sub-solar metallicity is derived from fitting only the \mgfef~index among the metal-sensitive indices (instead of \mgtwofe~and/or \mgfep). The lack of other Mg-Fe composite indices for the sub-solar-metallicity Q galaxies prevents a robust assessment of potential bias, although we note that their Fe4383 absorption index would indicate metallicities closer to or slightly above solar. Nevertheless, only 20\% of the Q galaxies fit with only \mgfef~have sub-solar metallicity. Excluding all the galaxies with only \mgfef~from the analysed sample would thus seem like an arbitrary choice. We rather keep these galaxies and only caution that the scatter of quiescent galaxies toward low metallicities at the high-mass end could be overestimated.  
In any case, the overall trend and fit to the mass-metallicity relation is not affected by these galaxies.
The star-forming galaxies display a steeper relation with larger scatter, and they are responsible for the change in slope around $10^{10.8}M_\odot$ observed in the general population (Fig.9 of Paper I). Star-forming galaxies reach stellar metallicities as high as those of quiescent galaxies, with median values consistent within $1\sigma$ with those of quiescent galaxies, only above $10^{11}M_\odot$. At lower masses they show a small but systematic offset to lower metallicities with respect to quiescent galaxies. A simple linear fit is not adequate for the star-forming mass--metallicity relation, and we choose to adopt the same sigmoidal function adopted for the global population (Equation 2 of Paper I):
\begin{equation}
\log \left<Z_\ast/Z_\odot\right> = \bar{P} + A\cdot\tanh\left(B\cdot\log\frac{M_\ast}{\bar{M_\ast}}\right) - C \label{eqn:met_mass}
\end{equation}
where $\bar{M_\ast}$~is the characteristic mass of inflection, A represents the dynamical range of the relation, B regulates the extent of the inflection range, and  and C is simply $\rm A\cdot tanh(B\cdot(11.5/\log M_\ast))$ so that $\bar{P}$ gives the zero-point of the metallicity at $10^{11.5}M_\odot$.  We report the fitted parameters in Table~\ref{Tab:functional_fit_zstar_Q_SF}. 

\begin{figure}
\centering
\includegraphics[width=0.5\textwidth]{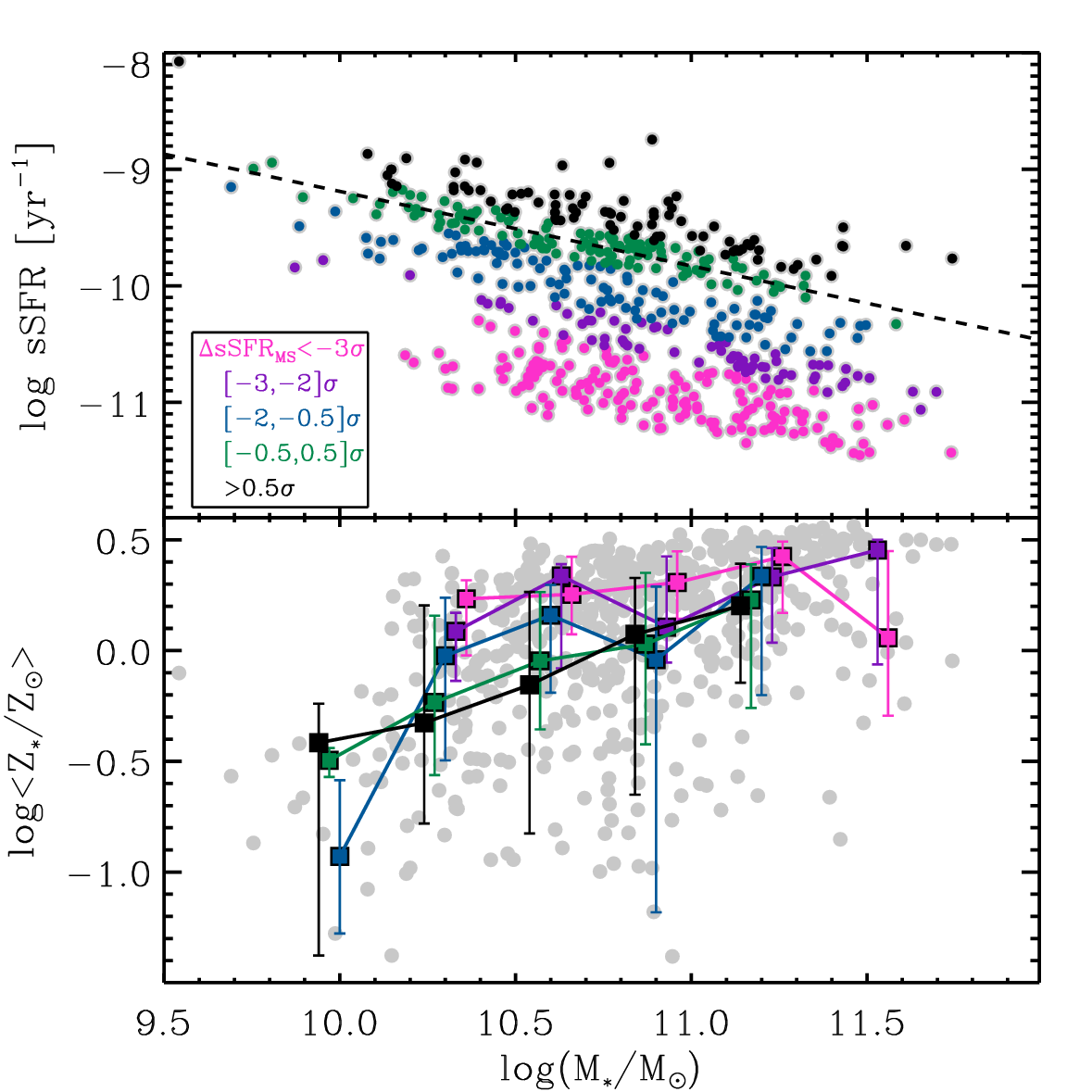}
\caption{Top panel: SSFR versus stellar mass for \silver~galaxies, divided into bins of distance from the relation fit to $UVJ$~star-forming galaxies (dashed line, Eq.~\ref{eqn:ssfr_mass}). Bottom panel: Luminosity-weighted mean stellar metallicity as a function of stellar mass for LEGA-C \silver~sample (grey points). Filled squares show the median stellar metallicity in bins of stellar mass (0.3-dex wide and with at least 5 galaxies), weighted by ${\tt Tcor}\times {\tt w\_spec\_silver}$, for galaxies in different bins of distance from the Main Sequence (as illustrated in the top panel). The errorbars represent the weighted $\rm16^{th}$~ and $\rm84^{th}$ percentiles. The unweighted median trends would be very similar to those shown here, but with narrower inter-percentile ranges, in particular for the intermediate-SSFR sub-sample.}
\label{fig:scaling_sfr_legac}
\end{figure}

Recently, \cite{Looser24} found the difference in stellar metallicity between quiescent and star-forming galaxies to result from a continuous secondary dependence of stellar metallicity on SFR. In Fig.\ref{fig:scaling_sfr_legac} we trace the median stellar metallicity as a function of stellar mass for \silver~galaxies separated into bins of distance from the Main Sequence in SSFR (Eq.~\ref{eqn:ssfr_mass}). 

As already discussed, quiescent galaxies (deviating by more than $2\sigma$~below the Main Sequence; magenta and violet points) populate only the high (super-solar) metallicity portion of the diagram with negligible dependence on SSFR and display a shallow relation with mass. Conversely, especially below $10^{11}\,\mathrm{M_\odot}$, galaxies closer to or on the Main Sequence are spread over a larger range in stellar metallicity (with increased downward scatter) than galaxies with lower SSFR, leading to a systematic shift toward lower average stellar metallicities with increasing SSFR, at fixed stellar mass. Galaxies with low stellar metallicity ($\log\left<Z_\ast/Z_\odot\right> <-0.5$) have the highest SSFR. 
These general trends are robust against different SFR estimates, as we discuss in Appendix~\ref{appendix:sys}.
This trend is indicative of a secondary dependence of the stellar metallicity on the SFR, at fixed mass, qualitatively similar to the Fundamental Metallicity Relation (FMR)  for the gas-phase metallicity \citep[e.g.][]{Mannucci10}.  However, assessing the significance of the correlation would require a proper treatment of the asymmetric uncertainties in the low-metallicity regime as well as a larger statistical sample, in order to disentangle the effects of measurement errors and sample variance.
 Interestingly, \cite{MunozLopez25} recently collected a sample of almost 400 galaxies at redshift $0.1<z<0.9$ from different MUSE surveys and found that galaxies with younger ages and a larger number of (recent) star formation episodes show a larger scatter toward low stellar metallicities, suggesting a picture in line with our results.

\subsection{Trends with stellar velocity dispersion}\label{sec:scaling_sigma}

\begin{figure*}
\centering
\includegraphics[width=0.8\textwidth]{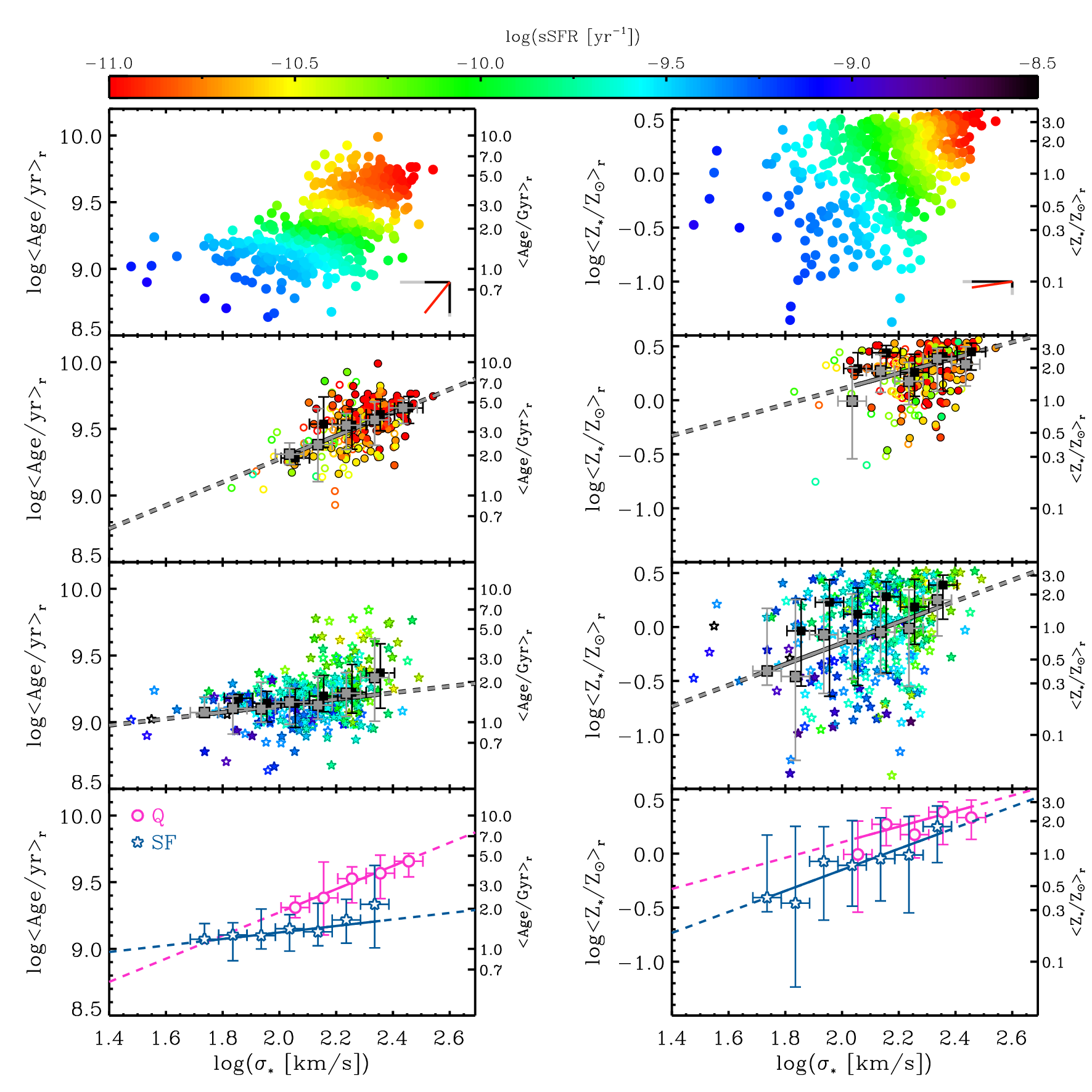}
\caption{Luminosity-weighted mean age (left panels) and mean stellar metallicity (right panels) as a function of stellar velocity dispersion for LEGA-C \silver~galaxies. Upper panels: the data points are color-coded by their SSFR after applying LOESS-smoothing. Black vectors in the bottom-right corner are proportional to the partial correlation coefficients computed as in \cite{ScholzDiaz24}, and the red vector indicates the direction of maximal increase of SSFR. Grey vectors indicate a reference correlation coefficient of 0.7. Middle panels: Quiescent galaxies (circles) and star-forming galaxies (stars), color-coded by their individual SSFR. Filled symbols highlight galaxies in the \golden~subsample. Median trends in bins of $\sigma_\ast$ (0.1~dex wide and with at least 5 galaxies) are shown by grey (black) squares for the weighted \silver~(non weighted \golden) sample. Linear fits to the median \silver~points are shown by solid lines (dashed lines for the extrapolation outside the median data points). Bottom panels: comparison of the median trends for quiescent (magenta) and star-forming galaxies (blue) in the \silver~sample.  The errorbars represent the 16-84 percentile range of each distribution. }
\label{fig:scaling_Q_SF_legac_sigma}
\end{figure*}

Figure~\ref{fig:scaling_Q_SF_legac_sigma} shows how age and stellar metallicity correlate with stellar velocity dispersion $\sigma_\ast$ in alternative to stellar mass (Fig.\ref{fig:scaling_Q_SF_legac}). Symbols and colors have the same meaning as in Fig.~\ref{fig:scaling_Q_SF_legac}. The medians and percetiles are reported in Table~\ref{Tab:median_relations_sigma}. We see qualitatively similar trends of increasing age and metallicity with velocity dispersion as observed with stellar mass, but a few differences are worth noticing. We see a clear gradient in SSFR along the age-$\sigma_\ast$~relation. The partial correlation analysis shows that SSFR has a stronger correlation with age ($\rho=-0.63\pm0.05$) than with velocity dispersion ($\rho=-0.35\pm0.07$). In the $Z_\ast-\sigma_\ast$~plane the SSFR varies along the median relation with a stronger dependence on $\sigma_\ast$~($\rho=-0.57\pm0.05$) than on metallicity ($\rho=-0.31\pm0.07$), especially at $\sigma_\ast>10^{2.2}\,\mathrm{km/s}$. The correlation between the SSFR and the stellar population parameters, age and metallicity, appears similarly strong when viewed at fixed stellar mass as at fixed velocity dispersion.

The (mild) increase in metallicity of quiescent galaxies with $\sigma_\ast$ is equivalent to that with $M_\ast$, corresponding to $\sim0.2$~dex increase in $\log Z_\ast$ per dex in $M_\ast$ or $\sigma_\ast$. The relation between the median stellar metallicty and velocity dispersion can be well described by a linear function, also for the star-forming galaxies. The linearly decreasing median metallicity with decreasing $\sigma_\ast$ is infact the result of an increasing scatter in stellar metallicity from high to low $\sigma_\ast$. We fit a linear function to the metallicity-velocity dispersion trends: 
\begin{equation}
    P = P_0 + \alpha \cdot \log(\sigma_\ast/10^{2.4} km/s) \label{eqn:met_sigma} 
\end{equation}
Star-forming galaxies follow a similar sequence as quiescent galaxies of decreasing metallicity and increasing scatter toward lower-mass/lower-$\sigma_\ast$~galaxies. They have similar slopes within the uncertainties, but the normalization is $\sim 0.15\pm0.08$~dex  lower for the SF galaxies. The best-fit parameters of the linear function are summarized in Table~\ref{Tab:rel_sigma}.

The light-weighted ages display a rather sharp transition from young ages of 1-2 Gyr to old ages of 3-5 Gyr around $\log\sigma_\ast\sim2.3$. We adopt the same linear function as in Eq.~\ref{eqn:met_sigma} to describe the age-velocity dispersion relation (see Table~\ref{Tab:rel_sigma}). The median age of star-forming galaxies increases only mildly with $\sigma_\ast$ below $\log\sigma_\ast\sim2.3$, while that of quiescent galaxies has a steeper  increase  with $\sigma_\ast$. 
 The zero-point of the relations for quiescent and star-forming galaxies differs by 2.5 Gyr ($0.4\pm0.04$~dex). 
However, there is a hint that that star-forming galaxies with the highest velocity dispersions have median ages only 1.5 Gyr ($0.24\pm0.09$~dex) younger than quiescent galaxies of similar $\sigma_\ast$. 
We checked that similar considerations apply, to a larger degree, to the mass-weighted ages. On the contrary, the two sequences of quiescent and star-forming galaxies remain well separated at fixed stellar mass also at the highest masses, with a difference of 3 Gyr ($0.41\pm0.14$~dex) between Q and SF at $M_\ast=10^{11.5}M_\odot$ (Fig.~\ref{fig:scaling_Q_SF_legac}). This suggests that the epoch when the bulk of the stellar population has formed depends primarily on the stellar velocity dispersion, regardless of whether galaxies are currently forming stars or not. 
The different star-formation histories and star formation efficiencies, however, lead to higher stellar metallicities in quiescent galaxies than in star-forming galaxies of similar velocity dispersion.

For the quiescent population, across the range $\rm \sigma_\ast=150-250~km/s$ we find a similar increase in stellar metallicity of $\sim0.1$~dex as found by \cite{Beverage23} and \cite{borghi22} (which is based though on the DR2 data reduction), and a slightly stronger increase in light-weighted age of 0.2~dex with respect to these studies.

\begin{table}
\centering
\caption{Fit to the median age-stellar mass relations in Fig.\ref{fig:scaling_Q_SF_legac} and~\ref{fig:legac_sdss_age}.}\label{Tab:functional_fit_age_Q_SF}
\begin{tabular}{|l|c|c|}
\hline
\multicolumn{3}{|c|}{$\log Age_r - \log M_\ast$ relation} \\
\hline
Sample & $P_0$ & $\alpha$ \\
\hline
LEGA-C \silver~Q & $9.69\pm0.01$ & $0.29\pm0.02$\\
LEGA-C \golden~Q & $9.67\pm0.01$& $0.26\pm0.04$\\
SDSS SN10 Q & $9.790\pm0.004$ & $0.166\pm0.005$\\
\hline
LEGA-C \silver~SF & $9.29\pm0.02$ & $0.14\pm0.02$ \\
LEGA-C \golden~SF & $9.29\pm0.02$ & $0.13\pm0.02$ \\
SDSS SN10 SF & $9.45\pm0.01$& $0.084\pm0.008$\\
\hline
\end{tabular}
\tablefoot{Relations fit to the median age as a function of stellar mass for Quiescent and Star-Forming galaxies, using a linear function (Eq.~\ref{eqn:age_mass}), for the \silver~and \golden~LEGA-C samples, displayed in the left panels of Fig.\ref{fig:scaling_Q_SF_legac}, and for SDSS DR7, shown in the middle and right panels of Figs.\ref{fig:legac_sdss_age}. The columns give: (1) sample, (2) intercept $P_0$ at $10^{11.5}M_\odot$, (3) slope $\alpha$. Trends for the \silver~sample are weighted for volume and completeness.}
\end{table}

\begin{table*}
\centering
\caption{Fit to the median stellar metallicity-stellar mass relations in Fig.~\ref{fig:scaling_Q_SF_legac} and~\ref{fig:legac_sdss_zstar}.}\label{Tab:functional_fit_zstar_Q_SF}
\begin{tabular}{|l|c|c|c|c|}
\hline
\multicolumn{5}{|c|}{$\log Z_{\ast,r} - \log M_\ast$ relation} \\
\hline
\noalign{\vspace{3pt}}

Sample &  $\bar{P}$ & A & B & $\bar{M_\ast} [M_\odot]$ \\
\hline
LEGA-C \silver~Q & $0.38\pm0.04$ & $0.6\pm5$ & $2\pm3$& $1\pm8\cdot10^{10}$ \\
LEGA-C \golden~Q & $0.5\pm0.1$ & $0.5\pm15$& $1\pm11$& $1\pm70\cdot10^{10}$\\
SDSS SN20 Q     & $0.437\pm0.002$& $0.4\pm0.1$& $0.6\pm0.1$ & $1.0\pm0.8\cdot10^{10}$\\
\hline
LEGA-C \silver~SF & $0.41\pm0.09$ & $1\pm7$& $1\pm4$ & $1\pm20\cdot 10^{10}$\\
LEGA-C \golden~SF & $0.44\pm0.01$& $1\pm2$& $1\pm1$ & $1\pm3\cdot 10^{10}$ \\
SDSS SN20 SF      & $0.40\pm0.01$& $0.6\pm0.1$ & $0.7\pm0.2$ & $1\pm0.4\cdot10^{10}$\\
\hline
\end{tabular}
\tablefoot{Relations fit to the median stellar metallicity as a function of stellar mass for Quiescent and Star-Forming galaxies, using the functional form of Eq.\ref{eqn:met_mass}, for the \silver~and \golden~LEGA-C samples, displayed in the right panels of Fig.\ref{fig:scaling_Q_SF_legac}, and for SDSS DR7, displayed in the middle and right panels of Figs.\ref{fig:legac_sdss_zstar} (for the $S/N>10$~and $S/N>20$~samples, respectively). The columns give: (1) sample, (2) characteristic parameter $\bar{P}$ at $10^{11.5}M_\odot$, (3) $A$~parameter, regulating the metallicity increase over the inflection, (4) $B$~parameter, regulating the mass range of inflection, (5) characteristic mass $\bar{M_\ast}$ of the inflection point. Trends for the \silver~sample are weighted for volume and completeness.}
\end{table*}

\begin{table*}
\caption{Fit to the median age and metallicity as a function of stellar mass for the general population.}\label{Tab:functional_fit_all}
\centering
\begin{tabular}{|l|l|c|c|c|c|}
\hline
\noalign{\vspace{3pt}}
parameter & sample  & $\bar{P}$ & A & B & $\bar{M_\ast} [M_\odot]$ \\
\hline
$\log<Age/yr>_r$ & LEGA-C \silver    & $9.67\pm0.08$  &   $0.3\pm0.2$  &  $2\pm1$ & $1.6\pm1.2 \cdot 10^{11}$ \\
                             & LEGA-C \golden & $9.66\pm0.02$  &   $0.30\pm0.02$  &   $2.4\pm0.3$ & $1.07\pm0.09\cdot 10^{11}$ \\
\cline{2-6}
                             & SDSS-SN10 & $9.782\pm0.002$  &   $0.289\pm0.005$  &   $1.28\pm0.04$ & $6.5\pm0.2\cdot 10^{10}$ \\
\hline
$\log<Z_\ast/Z_\odot>_r$ & LEGA-C \silver    & $0.38\pm0.03$ &  $0.9\pm1.6$ &   $1.5\pm1.2$ &  $1\pm2\cdot 10^{10}$ \\
                             & LEGA-C \golden & $0.46\pm0.01$ &   $1.0\pm1.7$ &    $1.2\pm0.6$ &  $0.9\pm2\cdot10^{10}$ \\
\cline{2-6}                             & SDSS-SN20 & $0.435\pm0.004$ &   $0.46\pm0.08$ &    $0.8\pm0.1$ &  $1.0\pm0.4\cdot 10^{10}$ \\\hline
\end{tabular}
\tablefoot{Relations fit to the median age and metallicity as a function of stellar mass, using the functional form of Eq.\ref{eqn:met_mass}, for the \silver~and \golden~LEGA-C samples as well as for SDSS, and displayed in the left panels of Fig.\ref{fig:legac_sdss_age} and Fig.\ref{fig:legac_sdss_zstar}. The columns give: (1) Parameter, (2) sample, (3) characteristic parameter $\bar{P}$ at $10^{11.5}M_\odot$, (4) \textit{A} parameter, regulating the age/metallicity increase over the inflection, (5) \textit{B} parameter, regulating the mass range of inflection, (6) characteristic mass $\bar{M_\ast}$ of the inflection point.}
\label{Tab:fit_relations}
\end{table*}

\begin{table}
\centering
\caption{Linear fit to age and stellar metallicity as a function of velocity dispersion.}\label{Tab:rel_sigma}
\begin{tabular}{|l|c|c|}
\hline \hline
\multicolumn{3}{|c|}{$\log Age_r - \log \sigma_\ast$ relation} \\
\hline
Sample & $P_0$ & $\alpha$ \\
\hline
LEGA-C \silver~Q & $9.62\pm0.02$ & $0.87\pm0.1$\\
LEGA-C \golden~Q & $9.65\pm0.02$& $1\pm0.1$\\
\hline
LEGA-C \silver~SF & $9.22\pm0.03$ & $0.24\pm0.1$ \\
LEGA-C \golden~SF & $9.23\pm0.04$ & $0.1\pm0.08$ \\
\hline \hline
\multicolumn{3}{|c|}{$\log Z_\ast - \log \sigma_\ast$ relation} \\
\hline
Sample & $P_0$ & $\alpha$ \\
\hline
LEGA-C \silver~Q & $0.39\pm0.05$ & $0.7\pm0.4$\\
LEGA-C \golden~Q & $0.43\pm0.04$& $0.30\pm0.2$\\
\hline
LEGA-C \silver~SF & $0.24\pm0.06$ & $0.97\pm0.15$ \\
LEGA-C \golden~SF & $0.39\pm0.04$ & $0.49\pm0.08$ \\
\hline
\end{tabular}
\tablefoot{Linear relations (Eq.~\ref{eqn:met_sigma}) fit to the median stellar age and stellar metallicity as a function of velocity dispersion for Quiescent and Star-Forming galaxies, for the \silver~and \golden~LEGA-C samples, displayed in the left panels of Fig.\ref{fig:scaling_Q_SF_legac_sigma}. The columns give: (1) sample, (2) intercept $P_0$ at $10^{2.4}\rm km/s$, (3) slope $\alpha$. Trends for the \silver~sample are weighted for volume and completeness.}
\end{table}

\section{The evolution of the scaling relations from $z=0.7$\ to $z=0.1$}\label{Sec:evolution}

Having established the stellar population scaling relations for the massive galaxy population at $z=0.7$, we now wish to quantify how much they have changed in the last 5 Gyr of cosmic history. To this end, we compare the relations obtained in LEGA-C with those of the local galaxy populations from SDSS DR7. 
\subsection{The SDSS DR7 data and sample}
For a proper comparison with our LEGA-C analysis, we take the revised estimates of stellar population parameters for SDSS DR7 that we derived in \cite{Mattolini25}, which adopt the same modeling assumptions and consistent observational constraints as in this work. Specifically, both LEGA-C and SDSS spectra are interpreted with the same library of SFHs, metallicity histories and dust attenuation parameters, based on the CB19 models. For SDSS we fit the $u,g,r,i,z$~photometry together with the optimal set of absorption features (\dn, \hb, \hdg, \mgfep, \mgtwofe). In addition to the model library adopted, there are other two novelties in the SDSS analysis with respect to our previous catalogs \citep{gallazzi05,Gallazzi21}: i) statistical weights including completeness and volume corrections and the treatment of duplicate spectroscopic observations, and ii) the corrections for aperture effects. Galaxies with more than one spectroscopic observation are included only once and their physical parameters are estimated from the S/N-weighted mean of the observed spectral indices (in addition to the unique photometry). More importantly, corrections for aperture effects are applied to the absorption indices of each spectrum. The effect of a fixed, finite aperture, sampling a redshift-dependent portion of the galaxy's light, on the measured absorption indices has been estimated in a statistical sense using simulations based on the CALIFA IFU spectroscopy \citep{Zibetti25}. These corrections were then applied to each SDSS galaxy, based on its redshift, measured index value in the fiber, rest-frame $g-r$ color, absolute $r$-band Petrosian magnitude and half-light radius, resulting in an estimate of the absorption index integrated over the full galaxy extent. These aperture corrections typically act in the sense of reducing the \dn\, while increasing the strength of Balmer-line indices (\hb, \hdg); the metal-sensitive indices (\mgfep, \mgtwofe) have typically negative corrections \citep{Zibetti25}. In terms of stellar populations, this leads in general to an overall decrease of both age and metallicity estimates, as expected from the predominant stellar population gradients observed in nearby galaxies. Notably, the strongest corrections apply to ``green valley'' galaxies, with a net effect of enhancing the young population at the expenses of the old one. On the metallicity side, aperture corrections produce a steepening of the mass-metallicity relation, especially at the low-mass end. A thorough analysis of these effects is presented in \cite{Mattolini25}.

The SDSS DR7 sample is limited to galaxies with good quality photometric flags, $0.005\leq z\leq0.22$, $0<V_\mathrm{DISP}<375 \,\mathrm{km/s}$\footnote{For galaxies with duplicate observations, redshift and velocity dispersion are the weighted mean of all the observations.}, $14.5\leq r_{Petro}\leq 17.77$. We further apply a cut in spectral S/N at $S/N>10$~and $S/N>20$~for age and stellar metallicity respectively, following \cite{Mattolini25}. The resulting samples of $354~977$ galaxies with $\mathrm{S/N}\geq10$, and $89~852$ galaxies with $\mathrm{S/N}\geq20$ are compared to the purely magnitude limited photometric catalog of \citep{blanton05} in the rest-frame color--absolute-magnitude plane, in order to derive statistical corrections for incompleteness induced by our spectroscopic selection. For both $\mathrm{S/N}$~limits, the sample is representative and statistically weighted to reproduce a volume complete selection down to $M_\star \sim 10^9\, \mathrm{M_\odot}$. 

In Fig.\ref{fig:legac_sdss_age} and \ref{fig:legac_sdss_zstar} we compare the age and stellar metallicity distributions as a function of stellar mass for LEGA-C galaxies with those for SDSS DR7 galaxies. The contours and the median trends for SDSS are computed weighing galaxies for volume and spectroscopic completeness, as in \cite{Mattolini25}. They should thus be regarded as representative of a volume complete sample, similarly to the median trends of the LEGA-C \silver~sample. In these figures we show the distributions and trends for the population as a whole (left panels) and for quiescent and star-forming galaxies separately (middle and right panels). LEGA-C galaxies are classified into quiescent and star-forming as described in Sec.~\ref{sec:data_sample}.  We follow an equivalent approach as for LEGA-C to classify Q and SF galaxies in SDSS: we consider as Q those galaxies that lie more the $2\sigma$~below the main sequence in the $\rm SSFR-M_\ast$~plane. Specifically, we use aperture-corrected SFR estimates based on the $H\alpha$~emission following \cite{Jarle04}. To define the MS we consider galaxies classified as star-forming based on the BPT diagram and we fit a linear relation between their SSFR and $M_\ast$ (see Eq. 7 in \citet{Mattolini25} and Fig. 6 in \citet{Gallazzi21}). In this way we classify galaxies into Q and SF depending on their relative SSFR with respect to the MS at each redshift. This criterion is also less sensitive to absolute differences in SFR estimates from different diagnostics (see also discussion in Appendix~\ref{appendix:sys}). 

\subsection{The evolution in the age-mass relation}
\begin{figure*}
    \centering
        \includegraphics[width=\textwidth]{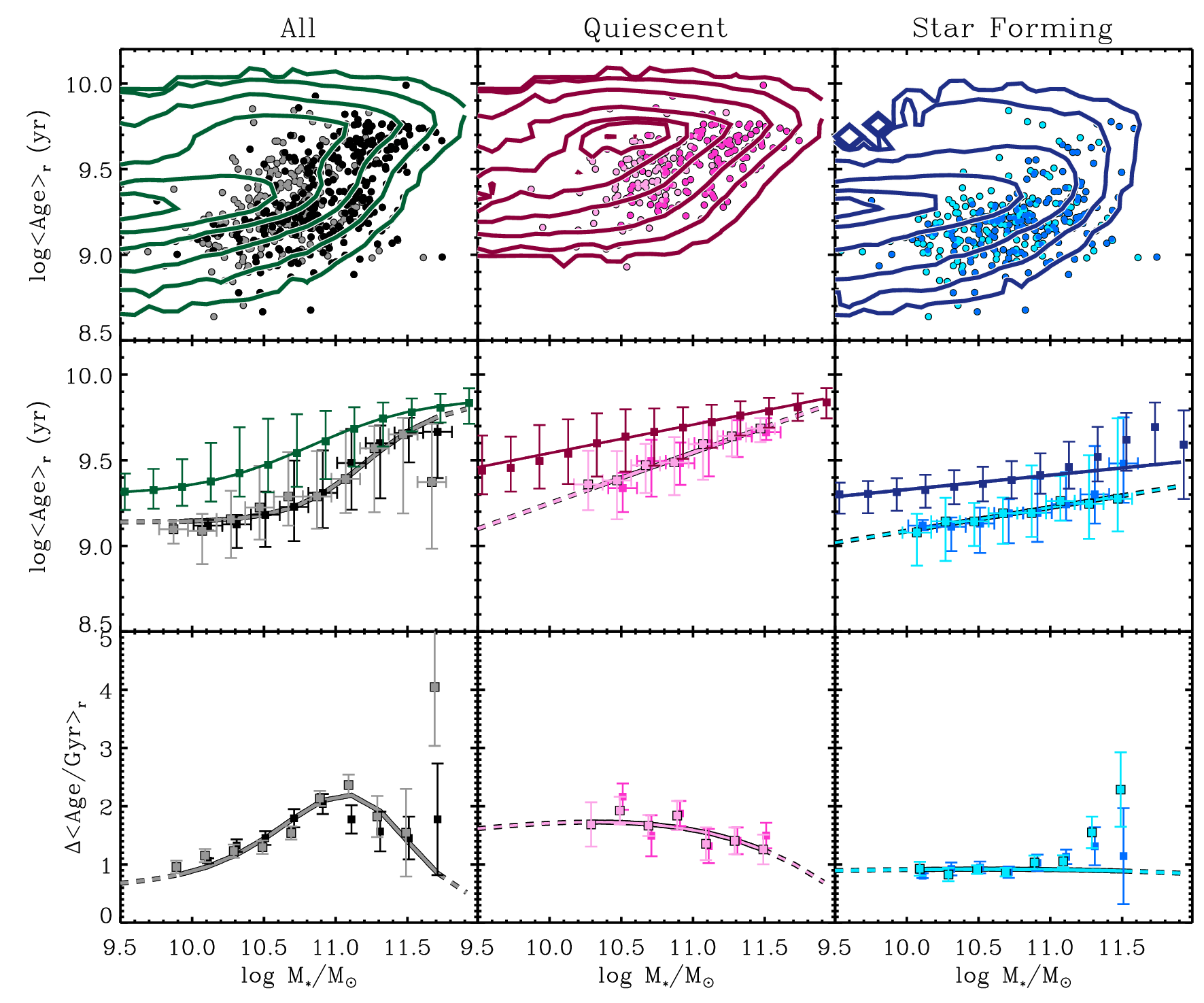}
        \caption{Comparison of the light-weighted age versus stellar mass relation for LEGA-C and SDSS galaxy samples, for the whole population (left panels) and for quiescent and star-forming galaxies separately (middle and right panels, respectively). The SDSS data include corrections for aperture effects, as well as weights for spectroscopic and volume completeness (see text for details). {\it Top row}: Data points show LEGA-C \golden~(black filled and dark color points) and \silver~samples (grey and light color points). The contours show the distribution for SDSS DR7 $S/N>10$~samples, tracing number density levels that enclose 16, 50, 84, 97.5, 99.8, 99.9\% of the total density. {\it Middle row:} median values of light-weighted age in bins of stellar mass for LEGA-C \golden~and \silver~sample (weighted for volume and spectroscopic completeness) and for SDSS, with the same color coding as in the top row. The errorbars indicate the (weighted) 16-84 percentiles of the distribution. The solid curves show the fit to the median \silver~points, with the same functional form adopted in Paper I for the whole population (as Eq.\ref{eqn:met_mass} for metallicity) and with a linear function (Eq.\ref{eqn:age_mass}) for Q and SF galaxies. The dashed lines show the extrapolation beyond the range of the data. {\it Bottom row:} Difference in (linear) age between SDSS median points and LEGA-C (\golden~and \silver) median points, with errorbars computed from the error on the medians; the solid lines show the differences between the fitted functions in the middle row. The y-axis range extends to 5~Gyr, which corresponds to the cosmic time elapsed between z=0.7 and z=0.1, i.e. roughly the expected age difference under pure passive evolution.}
        \label{fig:legac_sdss_age}
\end{figure*}

In Fig.\ref{fig:legac_sdss_age} we clearly see a shift to younger ages from SDSS to LEGA-C, as expected, when comparing the population as a whole (left panels) and quiescent galaxies (middle panels). The SDSS contours for the population as a whole show a bimodal age distribution with an old and a young sequence, transitioning around $3-6\cdot10^{10}M_\odot$. This is qualitatively similar to what we observe in LEGA-C at $z\sim0.7$. 
The SDSS distribution of light-weighted ages extends between 800~Myr and 10~Gyr. LEGA-C galaxies probe only the high-mass end because of the K-band selection. The old age envelope for LEGA-C  (both for the whole population and for quiescent galaxies only) reaches 6.3 Gyr, consistent with the Universe age at z=0.7 and almost 4 Gyr younger than the SDSS old envelope. The median ages as a function of stellar mass for the SDSS general population can be well described by the same sigmoidal function adopted for LEGA-C (Eq.~\ref{eqn:met_mass}, see also Paper I). Compared to LEGA-C, the SDSS median relation is shifted to older ages ( $0.11\pm0.08$~dex)  offset in the characteristic age at $10^{11.5}M_\odot$) and displays a shallower transition between young and old occurring across  a stellar mass of ${\bar{M_\ast}}=6.5\pm0.2\cdot10^{10}M_\odot$, which is 0.4~dex smaller than at $\left<z\right>=0.7$ (see Table~\ref{Tab:functional_fit_all}). The value of the SDSS transition mass defined by our functional fit is consistent within $1\sigma$  with the value of $10^{10.80\pm0.05} M_\odot$  obtained in \cite{Mattolini25} for the mass where the number densities of young and old galaxies are equal. 

The bottom left panel of Fig.~\ref{fig:legac_sdss_zstar} shows the difference in linear age between the median of galaxies at z=0.1 in SDSS and the median of galaxies at z=0.7 in LEGA-C, in bins of stellar mass. Remarkably, over the whole mass range probed, the difference in the median age of the galaxy populations is significantly lower than the elapsed cosmic time of $\sim5$~Gyr, assuming the average redshift $\left<z\right>=0.7$ for LEGA-C and the average redshift $\left<z\right>=0.1$ for SDSS. The age difference is largest ($\sim2.3$~Gyr) around a stellar mass of $10^{11}M_\odot$, and decreases to 1 Gyr at lower stellar masses and to 1.5 Gyr at higher stellar masses. While the overall galaxy distribution, and in particular the old age envelope, shifts to older ages almost consistently with passive aging, the median of the population evolves more mildly. 

We further distinguish the contribution to the population evolution from quiescent and star-forming galaxies (middle and right columns). As in Fig.\ref{fig:scaling_Q_SF_legac}, we adopt a linear function (Eq.\ref{eqn:age_mass}) to describe the median trends with mass. The best-fit linear relations for each sample are reported in Table~\ref{Tab:functional_fit_age_Q_SF}. The median ages of LEGA-C quiescent galaxies show a slightly steeper (by $0.12\pm0.02$)  increasing trend with stellar mass with respect to SDSS, leading to an apparent evolution ranging from 1.5 to 2 Gyr in the mass range between $3\cdot10^{11}$ and $3\cdot10^{10} M_\odot$. Perhaps surprisingly, the largest age difference manifests for least massive quiescent galaxies, rather than for the most massive ones. This age difference is in any case notably smaller than the cosmic aging, i.e. what would be expected in the case of simple passive evolution\footnote{An even larger difference of 6~Gyr would be expected in the light-weighted age for the passive evolution of composite stellar populations \citep{gallazzi14}.}. The light-weighted ages of star-forming galaxies at $z\sim0.7$ show the same increasing trend with stellar mass as for $z\sim0.1$ star-forming galaxies, with an almost constant offset of $\sim0.2$~dex ($\sim 1$~Gyr). This suggests that, over the mass range $10^{10}-10^{11}\mathrm{M_\odot}$, the formation of new stars and the aging of the old stellar components balance to a 1~Gyr apparent evolution. There is a hint of a larger age difference between LEGA-C and SDSS star-forming galaxies at the highest masses, which is driven by the deviation from the linear trend in SDSS. This could be caused by the older ages of the bulges in local massive star-forming galaxies.

We also notice that the age difference between LEGA-C star-forming galaxies and SDSS quiescent galaxies ranges between 2 and 4 Gyr from low to high masses,  i.e. close to the expected 5 Gyr difference in case of passive evolution, as illustrated in Fig.~\ref{fig:Delta_Q_SF_evolution}. A rapid quenching of the massive z=0.7 star-forming galaxies would  thus bring them onto the median age-mass relation of local quiescent galaxies.

\subsection{The evolution in the stellar metallicity-mass relation}
Fig.\ref{fig:legac_sdss_zstar} compares the distribution in light-weighted mean stellar metallicity as a function of stellar mass for LEGA-C and SDSS galaxies. Contrary to age, no bimodality is observed in the stellar metallicity distribution as a function of stellar mass neither in the SDSS nor in the LEGA-C samples. LEGA-C and SDSS galaxies occupy the same region in the stellar-metallicity--mass plane over the common mass range, though LEGA-C shows an apparently larger scatter toward low metallicities. The median stellar metallicity of SDSS galaxies is fully consistent with that of equally-massive LEGA-C \golden~galaxies for masses above $\sim10^{10.7}M_\odot$. However, considering the (weighted) \silver~sample a difference in the medians of $\lesssim0.1$~dex is detected ($0.12\pm0.03$~ dex difference between medians at $\log M_\ast\geq11$). A marked decrease in median metallicity from $\left<z\right>=0.1$ SDSS galaxies to $\left<z\right>=0.7$ LEGA-C galaxies is observed below $\log M_\ast=10.5$, with differences between the two populations of at least 0.2~dex. 

Considering only the quiescent galaxies, we see that the distribution and median trend of stellar metallicity with mass are highly  consistent between the two redshifts over the mass range $3\cdot10^{11}-3\cdot10^{10} M_\odot$. The possible steepening of the stellar metallicity-mass relation of LEGA-C quiescent galaxies below $10^{10.5}M_\odot$ is driven by only few points in the \silver~sample and may suffer from incompleteness in the LEGA-C quiescent sample at these masses because of the LEGA-C K-band selection  and should thus be regarded with caution. The similarity of the stellar metallicity distribution for SDSS and LEGA-C quiescent galaxies suggests that any addition of young stars to the population of quiescent galaxies that explains the small age evolution (Fig.~\ref{fig:legac_sdss_age}), through either rejuvenation episodes in already quiescent galaxies or minor mergers or newly quenched galaxies, occurs in such a way to maintain galaxies on the same mass--metallicity relation.
It is interesting to note that for the highest mass bin ($M_\ast\gtrsim10^{11.5}M_\odot$) the median metallicity of LEGA-C Q galaxies is $\sim0.22\pm0.1$~dex  lower than for equally massive SDSS Q galaxies. A similar fall in total metallicity was reported in \cite{Carnall22} from the analysis of the stacked spectrum of $\log M_\ast/M_\odot>10.8$, $1<z<1.3$ quiescent galaxies in VANDELS. This suggests that the most massive galaxies population change in both age and stellar metallicity. We caution though that the median metallicity in the highest-mass bin may be biased low by a few galaxies whose sub-solar metallicity estimate is based only on the \mgfef~index (see Sec.\ref{sec:scaling_mass}).

The right panels of Fig. \ref{fig:legac_sdss_zstar} show that, at both redshifts, the mass--stellar metallicity relation of star-forming galaxies reaches a similar characteristic metallicity at $M_\ast=10^{11.5}M_\odot$  as the quiescent population ($\log(Z_\ast/Z_\odot)\sim0.4$; see Table~\ref{Tab:functional_fit_zstar_Q_SF}). At lower masses the relation is steeper for the $\left<z\right>=0.7$ star-forming population than at $\left<z\right>=0.1$. This leads to an increase in the median stellar metallicity of star-forming galaxies at masses below $10^{10.8}M_\odot$ by 0.05 up to more than 0.2~dex, across this redshift interval. 
\begin{figure}
    \centering
    \includegraphics[width=1\linewidth]{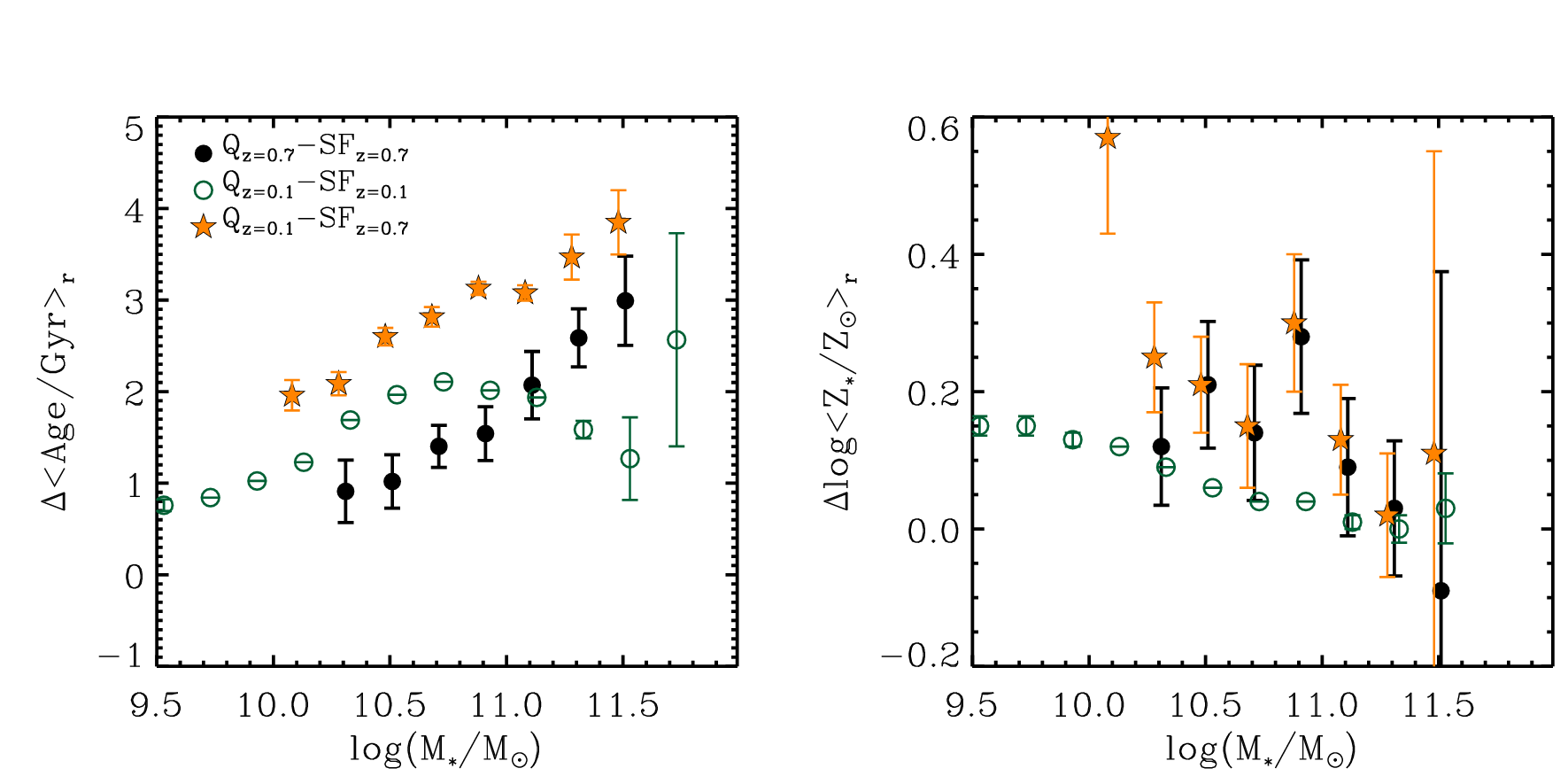}
    \caption{Difference in median light-weighted age (left panel) and median stellar metallicity (right panel) as a function of stellar mass between quiescent and star-forming galaxies at different redshifts. We compare Q and SF galaxies in LEGA-C (black circles) and in SDSS (green circles), as well as SF galaxies in LEGA-C with Q galaxies in SDSS, their potential descendants (orange stars). }
    \label{fig:Delta_Q_SF_evolution}
\end{figure}
Interestingly, the difference in median stellar metallicity as a function of mass between quiescent and star-forming galaxies is similar at both redshifts. It ranges between 0.03 and 0.2~dex from $10^{11.5}$ to $10^{10}M_\odot$ for SDSS, a behaviour that is quantitatively similar (consistent within 1$\sigma$), despite with larger uncertainties, for LEGA-C.  This is illustrated in Fig.~\ref{fig:Delta_Q_SF_evolution}.  The median difference in metallicity between quiescent and star-forming galaxies is thus not affected by the evolution of individual galaxies and of the population.

\begin{figure*}
    \centering
        \includegraphics[width=\textwidth]{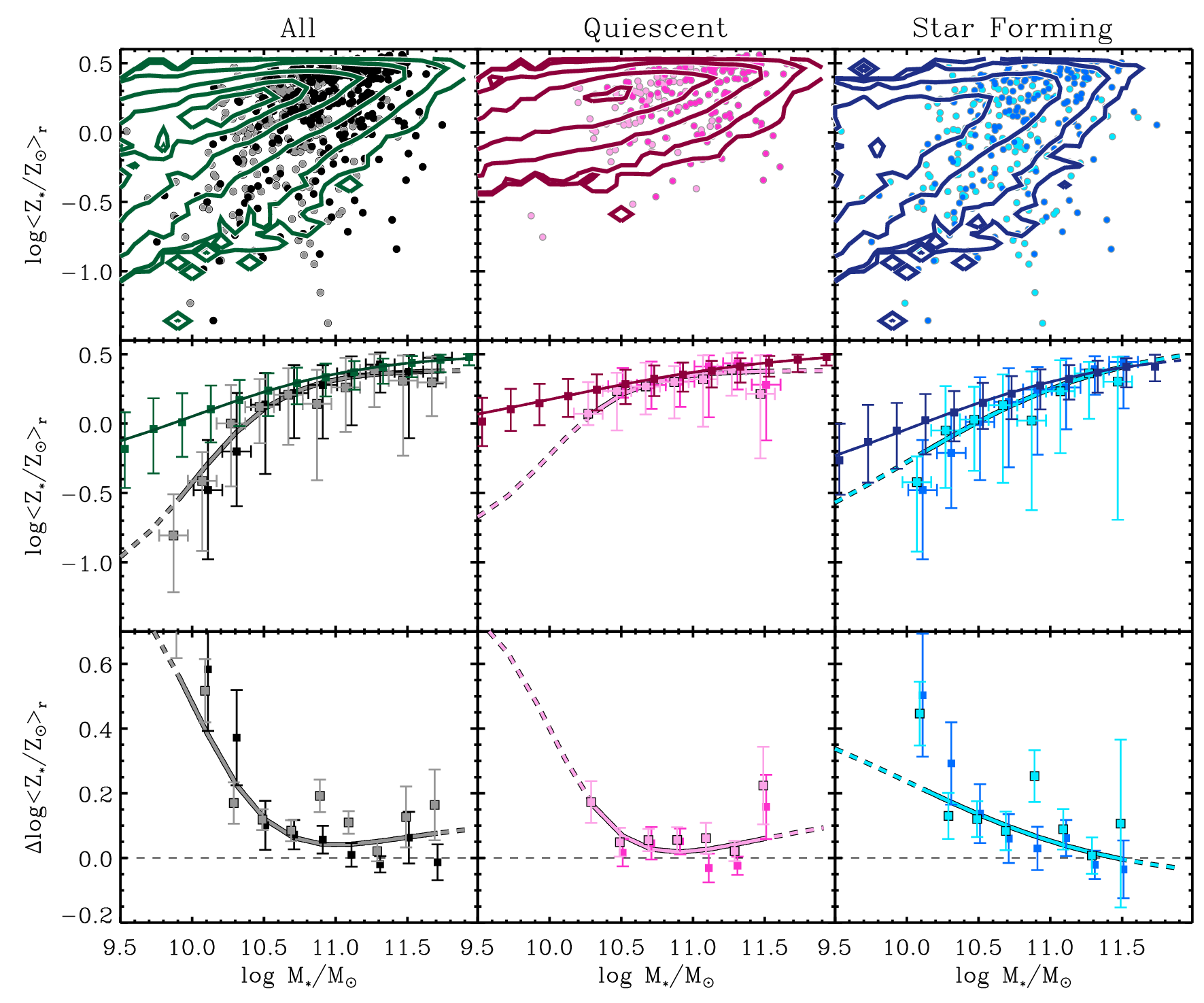}
        \caption{Comparison of the light-weighted stellar metallicity versus stellar mass relation for LEGA-C and SDSS galaxy samples, for the whole population (left panels) and for quiescent and star-forming galaxies separately (middle and right panels, respectively). Symbols, colors, and contours have the same meaning as in Fig.\ref{fig:legac_sdss_age}. The SDSS sample is restricted to $S/N>20$~galaxies and weights are applied to correct for volume and spectroscopic completeness. The bottom row shows the difference in (log) stellar metallicity between SDSS and LEGA-C median trends  with errorbars computed propagating the error on the medians. The dashed line at a difference equal to zero is drawn to guide the eye.}
        \label{fig:legac_sdss_zstar}
\end{figure*}

\subsection{Addressing systematics on the inferred evolution}
A few sources of systematics may limit the comparison between the stellar population scaling relations observed at different redshifts and the inferred evolution. We reduce at our best the potential biases by applying a consistent modeling and treatment of the data in the two samples. 

In particular, we account for observational biases due to aperture effects. We can assume that aperture effects are not significant for LEGA-C spectra, which are collected from $8"\times 1"$ slits with typical seeing comparable to the galaxy size in many cases. When the slit is aligned with the galaxy's major axis within $45\degree$~(which happens in $\sim50\%$ of the `primary' LEGA-C DR3 sample), the slit geometry allows to collect a good portion of the galaxy's light, with median slit loss of 45\% \citep{vanHoudt21}, compared to the 70\% of SDSS fiber loss. Moreover, the seeing smooths out gradients \citep{DEugenio20}, so that aperture corrections are not expected to be relevant for absorption features and inferred properties to within the typical uncertainties. Aperture effects can instead be significant for the SDSS fiber observations \citep[see][]{Zibetti25}. Because of stellar population gradients, accounting for aperture effects typically reduces the average galaxy ages.  The resulting scaling relations with and without corrections for aperture effects are qualitatively consistent, but differ in detail. In particular, accounting for aperture effects leads to an enhanced bimodality in the age-mass relation and a transition mass $0.15$~dex higher than without aperture corrections, and a steeper mass-metallicity relation at low masses \citep[these changes are discussed extensively in][and \cite{Zibetti25}]{Mattolini25}.  To quantify the impact of aperture effects, we checked the resulting evolution in the scaling relations that one would obtain  when analysing the original SDSS DR7 spectral indices without applying aperture corrections. The difference in light-weighted age between the SDSS and LEGA-C quiescent galaxy population  would increase  by $\sim1$ Gyr with respect to the aperture-corrected estimates, ranging between 2 and 3 Gyr from $10^{11}$ to $10^{10}M_\odot$. A reverse trend  would be  observed for star-forming galaxies, with the age difference between the two redshift going from 1 to 2 Gyr with increasing mass. Regarding stellar metallicity, we  would obtain  typically more metal-rich stellar populations at low masses, resulting in flatter scaling relations at $\left<z\right>=0.1$. We  would find  the median stellar metallicity of LEGA-C galaxies to be lower than their local counterparts already at masses of $10^{11}M_\odot$ and below, significantly for star-forming galaxies ($0.1\pm0.05$, i.e. $2\sigma$~above zero)  and marginally for quiescent galaxies (only $1\sigma$~above zero).

This shows that tracing stellar populations over consistent areas of the galaxies, hence including also more external regions than probed by the SDSS fiber, has important implications for the inferred evolution. The effect of including aperture corrections suggests that the addition of younger stellar components, which reduces the mean age evolution, pertains to a large extent to the outer galaxy regions, but does not explain alone the deviation from simple passive evolution.

In addition to aperture corrections, we checked what the effect on the inferred age evolution could be of degeneracies with dust attenuation in the age estimates of quiescent galaxies (see also Appendix C of Paper I). As a test, we performed a fit to both SDSS and LEGA-C restricting the model library to models with dust attenuation $A_g<0.2$~mag. Overall the Age-Mass relation of Q galaxies in LEGA-C shifts by 0.1 dex systematic, while that of SDSS Q galaxies becomes shallower and shifts by $<0.1$~dex at high masses and $\sim0.2$~dex at lower masses, leading to a difference in light-weighted age between SDSS and LEGA-C of $1.5-3\pm0.4$~Gyr in the mass range $10^{10.2}-10^{11.2}M_\odot$. The mass-metallicity relation of LEGA-C Q galaxies becomes flat, as a result of an increased scatter toward lower metallicities at all masses, while that of SDSS Q galaxies slightly steeper, leading to an apparent metallicity evolution of $0.15\pm0.05$~dex  at high masses $>10^{11}M_\odot$. Therefore, limiting the range of dust attenuations would alleviate (but still not remove) the deviation from passive aging, but at the same time would require metallicity evolution.

Finally, we repeat the comparison between SDSS and LEGA-C by adopting our previous model library as used in \cite{gallazzi05} and \cite{gallazzi14}. The old library adopted a constant metallicity along the SFH, an exponentially declining law for the smooth component of the SFH, and BC03 SSPs. The comparison thus incorporates differences both in SPS models and in SFH and metallicity assumptions. While the overall trends are robust against the different modelling, we find quantitative differences. We refer the reader to Appendix D of Paper I and to \cite{Mattolini25} for an in-depth presentation of the effect of each modeling ingredient on LEGA-C and SDSS, respectively. In particular, with the old library we find that the age-mass relation  for Q galaxies would become flatter and would move to slightly older ages for both SDSS and LEGA-C, implying an ages difference  between the two samples  of only $0.8\pm0.3$~Gyr  at the massive end and up to $2.8\pm1$~Gyr  at lower masses; the relation for SF galaxies would show a roughly mass-independent evolution of $1.3\pm0.3$~Gyr. The mass-metallicity relations would have a similar shape as in our default modeling  but shifted to lower metallicities by $0.2-0.3$~dex (due to changing from BC03 to CB19 models). The resulting   difference between SDSS and LEGA-C would be systematically higher by 0.2~dex, for both Q and SF galaxies. The indication of a non-passive evolution of the z=0.7 galaxy population accompanied by a mild stellar metallicity increase is thus a robust result, even assuming different priors (delayed-gaussian versus exponentially-declining continuous SFH; evolving versus constant stellar metallicity) and SPS models (CB19 versus BC03), at least in our modeling framework.  However, exactly how non-passive evolution occurs is subject to modeling assumptions and highlights the need of making consistent comparisons between different samples and redshifts. 

\section{Discussion}\label{sec:discussion}

\subsection{Different age sequences of quiescent and star-forming galaxies?}\label{sec:discussion_downsizing}
In Paper I we have shown that the light-weighted ages of z=0.7 galaxies follow a {\it non-linear} trend with stellar mass. In this work we show that this non-linear trend derives from the superposition of different distributions depending on the (S)SFR, so that quiescent and star-forming galaxies follow different distributions in the physical parameters space. Distinguishing galaxies into quiescent and star-forming, based on their SSFR, we find that they follow two separate {\it linear} trends in $\log\left<Age\right>_r-\log M_\ast$: a steeper relation for Q galaxies and a flatter one for SF, converging at relatively low masses (around the completeness limit of $10^{10.4}M_\odot$ for our Q galaxies sample). We observe a young and an old sequence also as a function of stellar velocity dispersion, with a rather sharp transition at $\log\sigma_\ast/\mathrm{km\,s^{-1}}\sim2.3$. This trend with velocity dispersion and the transition regime are very similar to those found for the full LEGA-C sample in \cite{Nersesian25} with ages estimated from {\tt Prospector}, and in \cite{Cappellari23} with {\tt pPXF}. \cite{Cappellari23} argues that the regime $\log\sigma_\ast/\mathrm{km\,s^{-1}}\sim2.3$ corresponds to a `quenching boundary' such that the star formation histories of galaxies above it have no young component (age$\rm\lesssim1\, Gyr$). By contrast, in our work, we find that the young and the old sequences as a function of velocity dispersion are not trivially associated to SF and Q galaxies: at $\log\sigma_\ast/\mathrm{km\,s^{-1}} > 2.3$ star-forming galaxies have similarly old ages as quiescent galaxies, with an age difference of 1.5~Gyr ($0.24\pm0.09$~dex), as opposed to how they compare at fixed $M_\ast$. {Similar considerations apply to mass-weighted ages: at fixed velocity dispersion star-forming and quiescent galaxies have more similar mass-weighted ages than they do at fixed stellar mass.} On the one hand, this suggests that stellar velocity dispersion is a better predictor of the main stellar formation epoch than stellar mass is, as we found in Paper I and as works in the local Universe also indicate \citep[e.g][]{vdW09,graves09a,Wake12,mcdermid15}. In this sense, our result is consistent with what is discussed in \cite{Cappellari23}. On the other hand, the fact that the old sequence contains both passive and star-forming galaxies, i.e. galaxies with different recent star formation histories, is in seeming contradiction with a `quenching boundary'.

\subsection{A redshift-independent stellar metallicity difference between quiescent and star-forming galaxies}\label{sec:discussion_metallicity}
 We find that Q and SF galaxies at $\left<z\right>=0.7$ reach the same stellar metallicity of $\log\left<Z_\ast/Z_\odot\right>_r\sim0.4$ at $10^{11.5}M_\odot$, but their median relations start to deviate at lower masses: the relation for Q galaxies is flat in the mass range $\log M_\ast/M_\odot=10.5-11.4$, while the relation for SF galaxies is steeper and with larger scatter. This leads to an average difference in metallicity of $0.11$~dex between Q and SF galaxies, with a trend of larger differences (of $\sim0.2$~dex or more) for galaxies with $\log (M_\ast/M_\odot)\leq11$. We verified that the difference in stellar metallicity between quiescent and star-forming galaxies and its trend with mass, is stable against changes in the modeling assumptions, namely in the SFH priors and SPS models. 
 However, we stress that there is no sharp separation between Q and SF galaxies in the stellar metallicity--mass relation. Above stellar masses of $10^{11}\,\mathrm{ M}_\odot$ there is significant overlap of galaxies with different SSFR, ranging from quiescent galaxies up to galaxies with $\log(\rm SSFR/\mathrm{yr}^{-1})=-9.5$. It is only for lower masses that a systematic shift toward lower  average  stellar metallicities with increasing SSFR is observed. The steepening of the stellar metallicity--mass relation has to be attributed in particular to galaxies that are on or above the star-forming main sequence (typically with $\log(\rm SSFR/\mathrm{yr}^{-1})>-10$). 
 We can translate the SSFR into a birthrate parameter as $b=\rm SSFR\cdot T\cdot(1-R)$ \citep[see][]{Gavazzi02,Jarle04}, which tells us how much the current star formation activity is with respect to the average past SFR, assuming for simplicity an average return mass fraction of $R=0.5$ and a time span T equal to the age of the Universe at $\left<z\right>=0.7$. Galaxies with $b<0.3$ follow a very similar stellar metallicity--mass relation. It is only for galaxies with $b\geq0.3$ that the stellar metallicity is lower than the one of equally-massive galaxies with a more quiescent star formation activity.

\cite{Bevacqua24} discusses the stellar metallicity--mass relation as the result of a lack of massive metal-poor galaxies among the quiescent population, rather than a simple increase in the average metallicity as a function of stellar mass. They introduce the concept of a MEtallicity-Mass Exclusion zone (MEME), whereby the stellar metallicity of massive quiescent galaxies is bounded between a constant upper limit and an increasing-with-mass lower limit. We find a similar asymptotic behaviour of the mass--metallicity relation toward high masses, as also found by other works in the gas-phase mass--metallicity relation. Notably in \cite{Bevacqua24} galaxies tend to accumulate at the upper boundary of their metallicity grid, an effect that is seen to a much less extent in our analysis, likely thanks to the larger range of metallicities in our adopted models. Most importantly, we do not find a Metallicity-Mass Exclusion zone for the quiescent population alone, which instead distributes along a relatively tight and flat sequence. A MEME-like relation as discussed in \cite{Bevacqua24} is instead possibly observed in the mass--metallicity relation of the population as a whole and in star-forming galaxies in particular, regardless of whether they are selected on the basis of SSFR or of UVJ location. In our analysis, such a behaviour is more evident in the distribution of stellar metallicity as a function of velocity dispersion rather than of stellar mass. It originates from an increasing scatter in metallicity toward lower mass/lower velocity dispersion. Our analysis shows that the SSFR contributes to this scatter.
 
This secondary dependence on SSFR resembles qualitatively the Fundamental Metallicity Relation that connects the gas-phase metallicity with the stellar mass and SFR, observed for star-forming galaxies \citep[e.g.][]{Mannucci10,Curti20}. 
The gas-phase metallicity is the end-point of the whole star formation and gas accretion histories, hence reflects the physical conditions of the gas coeval with the current star formation activity. Stellar metallicity instead is averaged over the whole SFH and traces metal enrichment at the earlier epochs of mass build-up. 
Therefore the physical origin of a dependence on SFR of the gas or the stellar component likely stems from mechanisms that operate on different timescales. An inverse relation between stellar metallicity and SSFR at given mass may be consistent either with substantial recent star formation fed by metal-poor gas inflows, or with less efficient metal recycling in galaxies with more extended and rising SFH.

A relation connecting the three parameters, stellar metallicity, stellar mass and (S)SFR, alike the one for the gas-phase metallicity, has also been predicted by semi-analythic models and cosmological hydrodynamical simulations. In particular the semi-analytic model GAEA, which includes a revised treatment of feedback from star formation, predicts a change in stellar mass--metallicity relation for galaxies with different SFR (different slope) or different SSFR (different normalization) \citep{fontanot21}. Similar predictions are obtained from cosmological hydrodynamical simulations such as Illustris, TNG and EAGLE, all featuring a treatment of the star-forming interstellar medium to simulate smooth stellar feedback \citep{Garcia24}. The dependence of stellar metallicity on star-formation rate is interpreted as a result of a correlation between the stellar and gas-phase metallicities and a balance between the timescales of ISM enrichment and of star formation. 

A systematic and mass-dependent difference in stellar metallicity between quiescent and star-forming galaxies at fixed stellar mass has been shown to exist in the local Universe first by \cite{peng15} using the stellar population catalog from \cite{gallazzi05}, and then confirmed by \cite{Trussler20} and by \cite{Gallazzi21}, using independent or updated estimates for SDSS DR7, and by \cite{Vaughan22} using IFU-integrated stellar population properties from SAMI. Moreover, a secondary dependence on SFR has been observed in star-forming galaxies by \cite{Looser24} based on MaNGA. The different degree of metal enrichment between quiescent and star-forming galaxies has been interpreted as suggestive of relatively long quenching timescales associated to gas starvation processes that suppress metal-poor gas supply to the galaxy \citep{peng15,Trussler20}. Other possible physical origins have been discussed, including a different efficiency of metal-loaded outflows in addition to starvation \citep{Trussler20} or in combination with different timescales of gas infall in present-day passive versus  star-forming galaxies \citep{spitoni17}. Moreover, galaxy structure as traced by mass density profile \citep{Zibetti22} and total potential well as traced by $\sigma_\ast$ or $M_\ast/R$ \citep{Barone22, Vaughan22, Looser24} appear to be important parameters to interpret the different stellar metallicities, possibly more fundamental than stellar mass, posing into question the necessity for long quenching timescales to reach high stellar metallicities.

In this work, we find that the secondary dependence of stellar metallicity on SFR is already in place at $\left<z\right>=0.7$. A small but systematic difference in stellar metallicity between quiescent and star-forming galaxies is observed either at fixed $M_\ast$, at low masses, or at fixed $\sigma_\ast$. This differs from what we observed for luminosity-weighted age (see section above), suggesting a different link to stellar and total mass (as traced by velocity dispersion) for star formation and chemical enrichment \citep{ScholzDiaz22,Barone22}. Interestingly, the observed difference in stellar metallicity and its trend with mass are quantitatively consistent within $1\sigma$  with what we find in the local Universe from SDSS \citep[see Fig.~\ref{fig:Delta_Q_SF_evolution}  and][]{Mattolini25}  and what found by the previously-cited works. This would suggest that its physical origin is independent of cosmic epoch at least over the last 5~Gyr.

\subsection{The non-simple passive evolution of massive galaxies in the last 5~Gyr}\label{sec:discussion_evolution}
We have traced the evolution of the age and stellar metallicity scaling relations in the last 5~Gyr, by comparing our results on LEGA-C galaxies at $\left<z\right>=0.7$ with those derived analysing SDSS galaxies at $\left<z\right>=0.1$ with the same modeling assumptions. We take advantage of the new characterization of the scaling relations in the local Universe from the analysis of \cite{Mattolini25} on SDSS DR7 spectra, which builds upon \cite{gallazzi05} and \cite{Gallazzi21} making important changes: i) spectral indices are corrected for aperture effects following \cite{Zibetti25}, and ii) completeness and Malmquist bias are accounted for with proper statistical weights. The LEGA-C and SDSS analysis are therefore consistent in the observational diagnostics used and in the modeling assumptions adopted, removing possible sources of bias in their comparison and the inferred evolution.

In case of fully passive evolution of galaxies, that is in absence of any addition of new stars either through star formation or through merging/accretion, we expect the stellar metallicity to remain constant. Indeed, after star formation ceases, the luminosity-weighted stellar metallicity that is inferred from the integrated galaxy spectra may decrease with the aging and fading of the younger stellar populations and the relative emergence of the older, metal-poorer stellar generations, in case of a general increase of the metallicity in galaxies along the SFH. This picture is in agreement  with the lack of evolution (within $1.5\sigma$), or even a very slightly negative evolution, in the median stellar metallicity between the quiescent galaxy populations at z=0.7 and at z=0.1 that we observe. The lack of significant changes in the total stellar metallicities and in the abundances of Mg and Fe of intermediate-redshift quiescent galaxies with respect to local quiescent galaxies is in agreement with results from \cite{Beverage23} \citep[which revises abundance estimates from][based on LEGA-C DR2 data]{Beverage21}, \cite{bevacqua23} and from \cite{saracco23} (extending to $z\sim1.4$ quiescent galaxies from VANDELS). Notice that this does not exclude merging in the massive galaxy population. In fact, because of the shallow slope of the stellar metallicity—mass relation for quiescent galaxies at masses above $10^{10.5}M_\odot$, mergers would leave the integrated metallicity constant within the uncertainties.

The median stellar metallicity of star-forming galaxies shows a modest increase between the z=0.7 population and the z=0.1 population at masses above $10^{10.8}M_\odot$. This is mostly associated to a reduction in the scatter toward low metallicities. Despite the large uncertainties, we find  tentative evidence that the scatter in the z=0.7 mass-metallicity relation is related to the galaxy SSFR, in the sense that more actively star-forming galaxies, at fixed mass, spread toward lower stellar metallicities 
(see Sec.\ref{sec:discussion_metallicity}). These galaxies are thus more prone to increase their metallicity over the 5 Gyr of cosmic evolution if their star formation continues at similar levels. At high masses, the gas-phase Fundamental Metallicity Relation flattens, so that the average gas-phase metallicity has little dependence on the (S)SFR \citep{Curti20, Mannucci10,Hunt12}, thus producing a constant increase in the stellar metallicity also for highly-star-forming galaxies. 
At lower stellar masses, we observe that the $z=0.7$ mass--metallicity relation of SF galaxies declines more steeply than the $z=0.1$ counterpart, leading to a median metallicity difference of $0.05$~dex up to more than $0.2$~dex in the mass range $\log(M_\ast/M_\odot)=10.8-10$. At first order approximation, this would imply a significant metal enrichment in the average star-forming population. Again, owing to the gas-phase FMR, which shows a stronger anticorrelation between gas-phase metallicity and SFR in this lower-mass regime with respect to higher masses\footnote{As shown in \cite{Curti20}, in the mass range $\log M_\ast=10.2-10.8$~and for galaxies on or above the Main Sequence, the gas-phase metallicity decreases by 0.3 dex for a 1.5 dex increase in SSFR at fixed mass.}, we expect  qualitatively that, while increasing their stellar mass, the more highly star-forming galaxies increase their stellar metallicity more slowly  than more quiescent systems, because new stars form out of less enriched ISM. Therefore, we expect that in the massive regime it is the more highly star-forming galaxies to contribute to the stellar metallicity evolution of the SF population, while at lower masses a larger contribution is expected from lower-SFR galaxies. A proper quantification of the expected increase in stellar metallicity requires simulations, which is beyond the scope of this paper.  

Along these lines, \cite{PS13} constrained the evolution in the stellar metallicity-mass relation of the present-day star-forming galaxies, assuming population-average SFH based on stellar mass and SFR estimates at different redshifts \citep{Leitner12} and the FMR \citep{Mannucci10}. At both $z=0$~and $z=0.7$, their predicted mass-metallicity relations (MZR) are similar to, but shallower than, our LEGA-C and SDSS observations,  so that their stellar metallicities overpredict our measurements at low stellar mass. Interestingly, the resulting evolution in the stellar MZR between z=0.7 and z=0 is consistent with our observed evolution in the median relation for masses $\log(M_\ast/M_\odot)>10.5$. Moving to lower masses the difference between the two redshifts increases more slowly with respect to our findings, reaching 0.2~dex at $10^{10}M_\odot$. This qualitative agreement is remarkable considering the different approaches. \cite{PS13} predictions apply strictly to present-day star-forming galaxies and to their progenitors, which may not be representative of the whole star-forming population at intermediate redshift, if a fraction of the z=0.7 star-forming population would quench and join the passive population today. Moreover, as mentioned above, our measured median difference for masses below $10^{10.3}M_\odot$ may be overestimated.

Predictions from the GAEA semi-analytic models \citep{fontanot21} also lead to a shallower stellar mass--metallicity relation for star-forming galaxies than what we observe. In agreement with our result, GAEA also predicts a mass-dependent evolution of the stellar metallicity, but more modest with respect to our findings. We find agreement for masses $>10^{10.7}M_\odot$, but at lower masses the GAEA semi-analytic model predicts a metallicity difference  between intermediate and low redshift  of only 0.1~dex at fixed mass. The low stellar mass regime is thus crucial to constrain different models.

Notice that the difference in the median stellar metallicity between LEGA-C SF galaxies and SDSS quiescent galaxies ranges between 0.1 and 0.3~dex, increasing with decreasing mass (Fig.~\ref{fig:Delta_Q_SF_evolution}). This difference is larger than the difference between LEGA-C SF galaxies and SDSS SF galaxies. This means that, if the whole SF population sampled by LEGA-C were to quench its star formation activity and join the Q population, it should experience an even larger metal enrichment than if it continued forming stars. Based on empirical predictions of the mass-metallicity relation of the star-forming progenitors of present-day passive galaxies, \cite{Trussler20} argued, through closed-box and leaky-box models, that these progenitors should go through an extended phase of starvation. Quenching timescales of $\sim 3$~Gyr are approximately compatible with a median difference in age of $\sim2.5$~Gyr between $z=0.7$~SF galaxies and local Q galaxies, as we observe. 
We should be cautious though in over-interpreting the metallicity trends and differences at masses below $10^{10.3}M_\odot$, which is the LEGA-C mass completeness limit at z=0.6 \citep{DR3}. In fact, below this mass  LEGA-C  may be missing $U-V$ red galaxies, and we should regard the mass--metallicity relation of SF galaxies at these masses as representative only of the $U-V<1.5$ galaxies.

In conclusion, the stellar metallicities of $z=0.7$~massive galaxies are overall compatible with passive evolution of the quiescent population and with the average cosmic SFH and gas-phase metallicity evolution for the star-forming population. The age distribution shows that a fraction of z=0.7 quiescent galaxies should evolve passively to reach the old boundary of the z=0.1 quiescent population. However, a pure passive evolution of the quiescent population is not reflected in the evolution of mean stellar age—stellar mass relation, which we observe to be significantly lower than the expected elapsed time between $\left<z\right>=0.7$ and $\left<z\right>=0.1$. Though quantitatively different, our analysis supports, with larger statistics and improved mass coverage, what we found in \cite{gallazzi14}. 
How can these two apparently contradicting results on age and metallicity evolution be reconciled?

An important point of our SDSS analysis is the correction for aperture effects. We estimate that the overall effect of accounting for total age and metallicity estimated through consistent physical apertures in SDSS is to reduce the apparent age evolution by 1~Gyr. This suggests that the addition of younger stellar components, through continued or revived star-formation or through minor merging/accretion, pertains to some extent to the outer regions of passive galaxies and of massive galaxies in general. Minor dissipationless mergers are considered an important channel of galaxy growth after quenching as inferred from the evolution of the mass-size relation \citep{Bezanson09,vdW09, Belli15}, the rotational support \citep{Bezanson18} and orbital structure \citep{DEugenio23b} of quiescent galaxies. However, this is not enough and additional evolution in the central regions of individual galaxies may be allowed. Small frosting or rejuvenation of star formation could be accomodated as it is not expected to alter the measured (mass-weighted) stellar metallicity \citep{ST07,saracco23}. Similarly, merging of similar-mass galaxies would move galaxies along the stellar mass axis but not on the stellar metallicity axis.

Moreover, we should remind that comparing scaling relations at different redshifts does not mean tracing the evolutionary tracks of individual galaxies.  Scaling relations are statistical “demographic” distributions, whose evolution is determined by the flow of galaxies in the multi-dimensional parameter space. The observed evolution in the scaling relations is the combination of the change in mass and physical properties of individual galaxies and the change in the number density of the population as a function of mass and SFR, including galaxies that fall off the K-band selection (hence mass completeness limit) of LEGA-C. The number and mass density of the global massive galaxy population has seen a modest increase since $z\sim1$~\citep{pannella09,moustakas13,leja20}. However, this seemingly small evolution hides a differential evolution of the quiescent and star-forming populations. Studies find the number density and mass density of massive quiescent galaxies to increase by 20-50\% (depending on mass threshold and analysis) since $z\sim1$~to the present \citep{moustakas13, Muzzin13b,diaz-garcia24}, while that of massive star-forming galaxies has remained constant \citep{Muzzin13b} or decreased \citep{moustakas13,Haines17}. This is also supported by the decrease of the mass range where we observed the age--mass relation to transition from being dominated by old quiescent galaxies to being dominated by young star-forming galaxies (0.4~dex smaller in SDSS than in LEGA-C). Our results suggest that quenching of star-forming galaxies over the last 6-8 Gyr contributes to the build-up of the massive quiescent population even at masses above $10^{11}M_\odot$.

We find that the massive SF galaxies in LEGA-C have a larger spread in stellar metallicities than quiescent galaxies, but on average they have metallicities as high as the quiescent population both at $z=0.7$~and at $z=0.1$. In particular, the high-metallicity SF galaxies are those with the lowest SFR at fixed mass. They are thus good candidates for joining the local quiescent population by exhausting their metal-rich (as expected from the FMR) ISM.\\

\section{Summary and conclusions}\label{sec:summary}
We analyse volume- and completeness weighted age and stellar metallicity scaling relations for Q and SF galaxies at redshift 0.6<z<0.77 from the LEGA-C survey. We use stellar population parameter estimates derived from our {\tt BaSta} fitting code that interprets in a Bayesian framework optimally selected sets of stellar absorption features, in combination with rest-frame optical photometry, with a library of spectral models based on complex star formation and metal enrichment histories, and dust attenuations (presented in Paper I).  We select a \silver~subset of the LEGA-C sample, based on the availability of key absorption features sensitive to both age and metallicity, comprising 232 quiescent and 320  star-forming galaxies, and a high-S/N \golden~subset, comprising 148 quiescent and 175  star-forming galaxies, where quiescent and star-forming are defined based on their distance from the $\rm SFR-M_\ast$ Main Sequence. We study how intermediate-redshift galaxies differentiate in stellar populations physical parameters according to their current star formation activity, and we find that:
\begin{itemize}
    \item Quiescent and star-forming galaxies follow two different sequences of age as a function of stellar mass, while their median ages are more similar if viewed at fixed velocity dispersion;
    \item Quiescent galaxies follow a shallow mass--metallicity relation above $M_\ast=10^{10.5}M_\odot$. Star-forming galaxies have similar metallicities as quiescent galaxies above $10^{11}M_\odot$, but are typically less metal-rich at lower masses. Similar differences are observed as a function of velocity dispersion;
    \item At masses lower than $10^{11}M_\odot$ the mass--metallicity relation of star-forming galaxies becomes steeper, with an increased scatter, which is associated with  the SSFR (relative to the SF Main Sequence), so that galaxies with higher SFR with respect to their past have on average lower stellar metallicity at fixed mass.    
\end{itemize}

We further compare our LEGA-C results with the volume-weighted and aperture-corrected scaling relations obtained through a consistent analysis of SDSS DR7. 
We find both similarities and differences in the physical parameters distribution of the massive galaxy populations at the two epochs:
\begin{itemize}
    \item We find a bimodal distribution in light-weighted mean age at both redshifts which originates from the different age--mass relations of quiescent and star-forming galaxies and the varying abundance of the two populations as a function of mass.
    \item The transition mass in the age--mass relations increases by 0.4~dex from $\left<z\right>=0.1$ to $\left<z\right>=0.7$. Despite the different definitions, this is similar to the evolution in the mass where passive and star-forming galaxies are equally dominant \citep{Muzzin13a,Haines17}. 
    \item While the upper age envelope of the local population (as a whole and for quiescent galaxies only) is $\sim4$~Gyr older than for LEGA-C, the median SDSS age--mass relation is shifted to older ages by only 1 to 2~Gyr for the star-forming and the quiescent population respectively.
    \item The stellar metallicity--mass relation shows negligible evolution between the two redshifts for masses above $10^{11}M_\odot$ and for quiescent galaxies at all masses probed. 
    \item For the star-forming galaxy population, we observe a steepening of the mass--metallicity relation starting at masses lower than $10^{11}M_\odot$~downwards, and possibly an increased scatter toward lower metallicities at all masses in LEGA-C with respect to SDSS. At face value this implies a metallicity evolution of less than 0.05~dex for masses above $10^{10.8}M_\odot$ and possibly exceeding 0.2~dex for lower masses down to $10^{10}M_\odot$.
    \item The relative differences between the mass--metallicity relation of quiescent and star-forming galaxies are quantitatively similar at $z=0.7$~and $z=0.1$.
\end{itemize}

These results suggest that the stellar metallicities of massive galaxies are set at early times. Combining secondary trends of SFR and expectations from the gas-phase FMR, the continued recycling of metals in star-forming galaxies is expected to impact the final stellar  metallicity in a different way for massive and less massive galaxies. 
Our analysis shows that the combined measurements of ages and stellar metallicities of both the quiescent and the star-forming populations can provide important constraints on the quenching and continued build-up of the massive galaxy population even in the more quiet Universe over the last 6 Gyr. The observed evolution in the scaling relations supports a scenario in which both individual evolution through rejuvenation or merging and population evolution through quenching  need to occur. 

Detailed comparison with model predictions and with the observed evolution in galaxy number densities and SFR can shed light on the relative contribution of continued star formation and of progenitor bias. We should also remind that the ability of stellar population models to resolve the early SFH is progressively more limited in older systems \citep{Zibetti24}. It is thus crucial to connect the local estimates and those at intermediate redshift in a continuous way, such as it will be possible with surveys as WEAVE-StePS \citep{weave-steps} and 4MOST-StePS \citep{4most-steps}, on one hand, and to constrain the ages and abundances of large samples of galaxies at increasing redshifts ($z>1$), for which high-sensitivity spectrographs in the NIR are essential, such as NIRSpec@JWST \citep[e.g. the SUSPENSE program,][]{Beverage25,Slob24}, PFS@Subaru \citep{PFS-GE} and MOONS@VLT \citep[the MOONRISE survey,][]{moonrise}, MOSAIC@ELT \citep{Mosaic_wp} in the near future. While the amount and quality of data at higher and higher redshifts increases, we stress that applying a consistent analysis in terms of modeling assumptions and observational constraints is paramount for a proper comparison of results at different redshifts.

\begin{acknowledgements}
We thank the referee for their detailed and constructive report that helped improve the presentation of the analysis and results. 
We thank Fabio Fontanot for discussion on the GAEA model predictions for the mass-metallicity relation.\\
ARG and LSD acknowledge support from the INAF-Minigrant-2022 "LEGA-C" 1.05.12.04.01.  SZ acknowledges support from the INAF-Minigrant-2023 "Enabling the study of galaxy evolution through unresolved stellar population analysis" 1.05.23.04.01. PFW acknowledges funding through the National Science and Technology Council grants 113-2112-M-002-027-MY2. LSD is supported by the "Prometeus" project PID2021-123313NA-I00 of
MICIN/AEI/10.13039/501100011033/FEDER, UE. This paper and related research have been conducted during and with the support of the Italian national inter-university PhD programme in Space Science and Technology.
\end{acknowledgements}
\bibliographystyle{aa}
\bibliography{paper}

\begin{appendix}
\onecolumn
\section{Robustness of quiescent and star-forming trends against different SFR and stellar population parameter estimates}\label{appendix:sys}
Here we address how much the comparison between quiescent and star-forming galaxies and the dependence on SSFR are robust against: i) the classification criterion between Q and SF and the SFR indicator; ii) the stellar population parameters estimates.

In Paper I we compared how a classification into Q/SF galaxies based on the SSFR (our choice) differ from one based on the $U-V, V-J$ colors (Fig.2 in Paper I). We noticed that these classifications are to a large extent consistent, with large overlap in the respective Q/SF samples. However, a classification based on $UVJ$ results in a larger sample of Q galaxies, thus including galaxies with SSFR closer to the Main Sequence. Our default selection based on SSFR is instead more conservative, classifying as quiescent the reddest galaxies in the $UVJ$~quiescent zone. 

\begin{figure}
    \centering
    \includegraphics[width=1\linewidth]{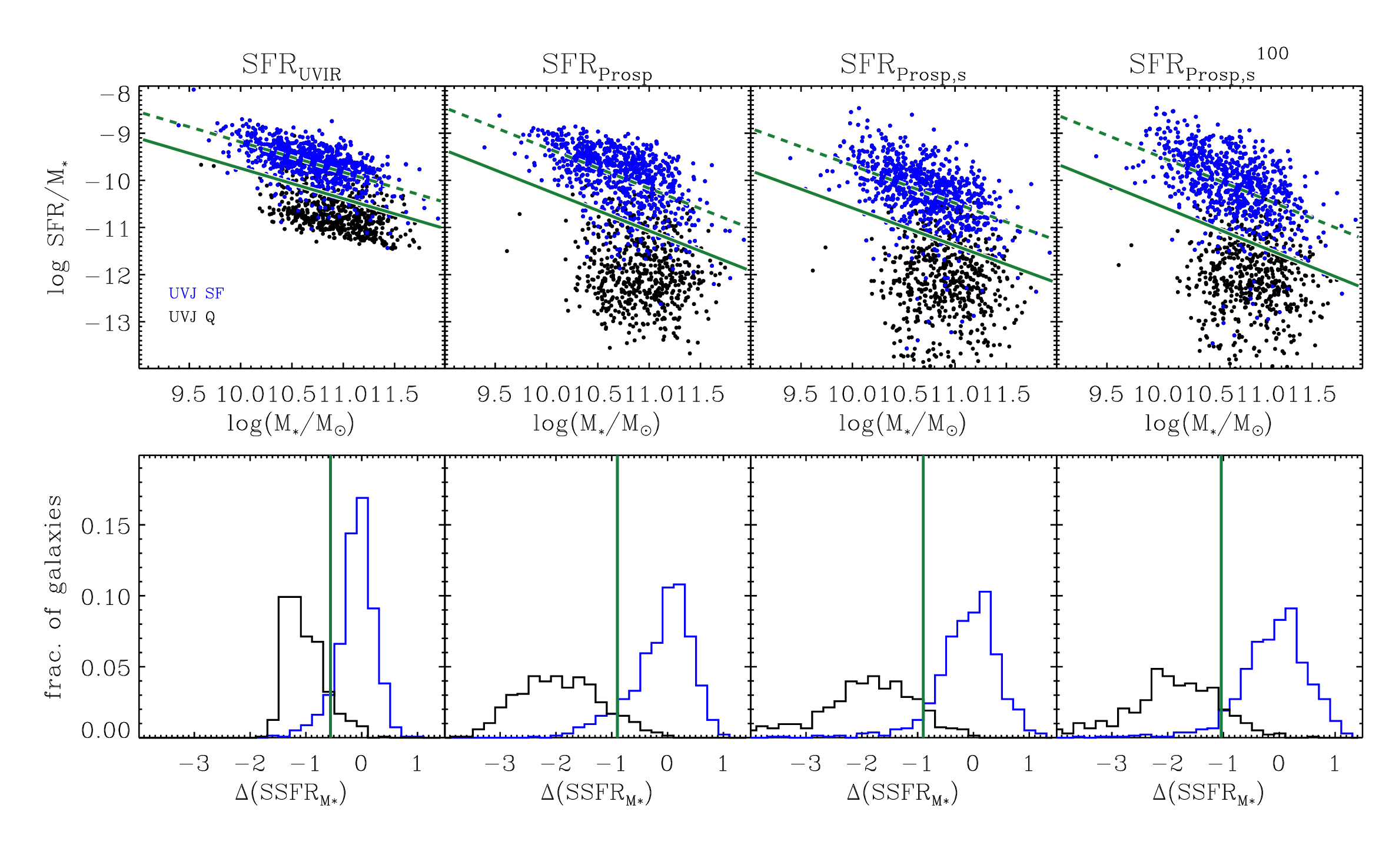}
    \caption{Comparison of Q/SF selection based on different SFR indicators. Galaxies are classified as Q if they lie 2$\sigma$~below the $\rm SSFR-M_\ast$~relation defined by UVJ-selected SF galaxies (blue points). The upper panels display the $\rm SSFR-M_\ast$~relation for different SFR estimates: our default choice based on UV+24\micron~($\rm SFR_{UVIR}$), estimate from Prospector fit to photometric data ($\rm SFR_{Prosp}$), from Prospector fit to spectrum+photometry instantaneous ($\rm SFR_{Prosp,s}$) or averaged over 100 Myr ($\rm SFR_{Prosp,s}^{100}$). Dashed and solid green lines indicate the fitted linear relation to UVJ-SF galaxies and its $2\sigma$~offset, respectively. The bottom panels show the histograms of distance from the Main Sequence for UVJ-SF (blue) and UVJ-Q (black) galaxies, with the selection cut indicated by the vertical green line. In this plot we show all LEGA-C galaxies in the redshift range of the \silver~sample.}
    \label{fig:compare_ssfr_selection}
\end{figure}

As default, we use SFR estimated from the UV and 24\micron~luminosities following \cite{Whitaker14} and \cite{bell05} ($\rm SFR_\mathrm{UVIR}$). In principle, the UV and IR (from monochromatic 24\micron~flux) luminosities trace the luminosity of young stars (direct or re-emitted by dust) and are thus good tracers of the star formation rate over timescales of $\lesssim100$~Myr \citep{Kennicutt98,bell05,bell07}. However, they are calibrated with standard assumption of constant SFR in the last 100~Myr and may be biased, especially for low SFRs, because of contribution to the 24\micron~flux from dust heated by processes not associated to star formation, such as old stars and/or AGN. On the other hand, SFR estimates based on SED fitting naturally account for the variety and complexity of SFHs and hence for different contributions from old stars, but inherit biases associated to degeneracies in the SFH and dust parameters \citep[see discussion in][]{Leja22}. For these reasons, we checked how much a criterion based on SSFR is sensitive to the adopted estimates of SFR. In particular, we considered other SFR estimates obtained from SED or spectral fitting with {\tt Prospector} as in \cite{Nersesian25}: i) instantaneous SFR from a fit to photometry only ($\rm SSFR_\mathrm{Prosp}$); ii) instantaneous SFR from a fit to spectroscopy plus photometry ($\rm SSFR_\mathrm{Prosp,s}$); iii) SFR averaged over the last 100 Myr from a fit to spectroscopy plus photometry ($\rm SSFR^{100}_\mathrm{Prosp,s}$).  We find a general agreement between $\rm SFR_{UVIR}$ and any of $\rm SFR_{Prosp}$, $\rm SFR_{Prosp,sp}$, $\rm SFR_{Prosp,sp}^{100}$ for highly star-forming (or UVJ-selected SF galaxies), with a scatter of 0.35, 0.6, 0.5~dex, respectively, although with a systematic offset of 0.3, 0.6, 0.5 dex toward higher $\rm SFR_{UVIR}$~with respect to the other estimates. Quiescent galaxies deviate to lower Prospector SFR estimates, with a scatter of 0.5-0.8 dex and systematic deviations ranging from 0.8 to 1.5 dex moving to lower-SFR galaxies. 
The slope and scatter of the relation between SSFR and M$_\ast$, as well as the spread in SFR of galaxies below the Main Sequence, are sensitive to the adopted SFR estimates, as illustrated in Fig.~\ref{fig:compare_ssfr_selection}. 
Nevertheless, UVJ-classified Q and SF galaxies separate well in $\rm SSFR-M_\ast$ for each of these SFR estimates. 
We consider all the LEGA-C galaxies within the redshift range of our \silver~sample and we fit a linear relation between SSFR and $M_\ast$ for each SFR indicator to the $UVJ$-classified star-forming galaxies. We then define those galaxies lying more than $2\sigma$ below the Main Sequence as quiescent (as indicated by the dashed and solid green line in each panel of Fig.~\ref{fig:compare_ssfr_selection}). This criterion identifies well the position where the quiescent and star-forming distributions cross for each SFR indicator, and results in consistent sample selection regardless of the SFR estimates adopted.  In Table~\ref{tab:numbers_sfr_selection} we compare the selection based on our default choice of $\rm SFR_{UVIR}$ with each of the other SFR estimates, as well as with the UVJ classification. For each pair, we report the number of galaxies that are classified as Q/SF by both methods and those with mismatched classifications. There is in general a $\sim$90\% agreement and a $\sim$10\% contamination. We note that the UVJ classification of Q galaxies includes low-SFR galaxies. We have checked that results in Fig.~\ref{fig:scaling_Q_SF_legac}~and~\ref{fig:scaling_Q_SF_legac_sigma} are not affected by differences in individual SFR estimates, as expected from the general good agreement on the Q/SF classification.   

\begin{table}
    \centering
    \begin{tabular}{|cc|cc|cc|cc|cc|}
     & & \multicolumn{2}{|c|}{$\rm SFR_{Prosp}$} & \multicolumn{2}{|c|}{$\rm SFR_{Prosp,s}$} & \multicolumn{2}{|c|}{$\rm SFR_{Prosp,s}^{100}$} & \multicolumn{2}{|c|}{UVJ}\\
     \hline
     & & Q & SF & Q & SF & Q & SF & Q & SF \\
     \hline
 $\rm SFR_{UVIR}$ & Q & 211 & 21 & 203 & 29  & 200 & 30  & 208 & 24 \\
                  & SF & 32 & 286 & 27  & 293 & 24  & 288 & 42  & 278\\
      \hline            
    \end{tabular}
    \caption{Confusion matrices comparing the Q/SF selection for the \silver~sample based on our default SSFR estimate ($\rm SSFR_{UVIR}$) with that based on other SSFR estimates or with UVJ classification (see text).}
    \label{tab:numbers_sfr_selection}
\end{table}

\begin{figure}
    \centering
    \includegraphics[width=0.4\linewidth]{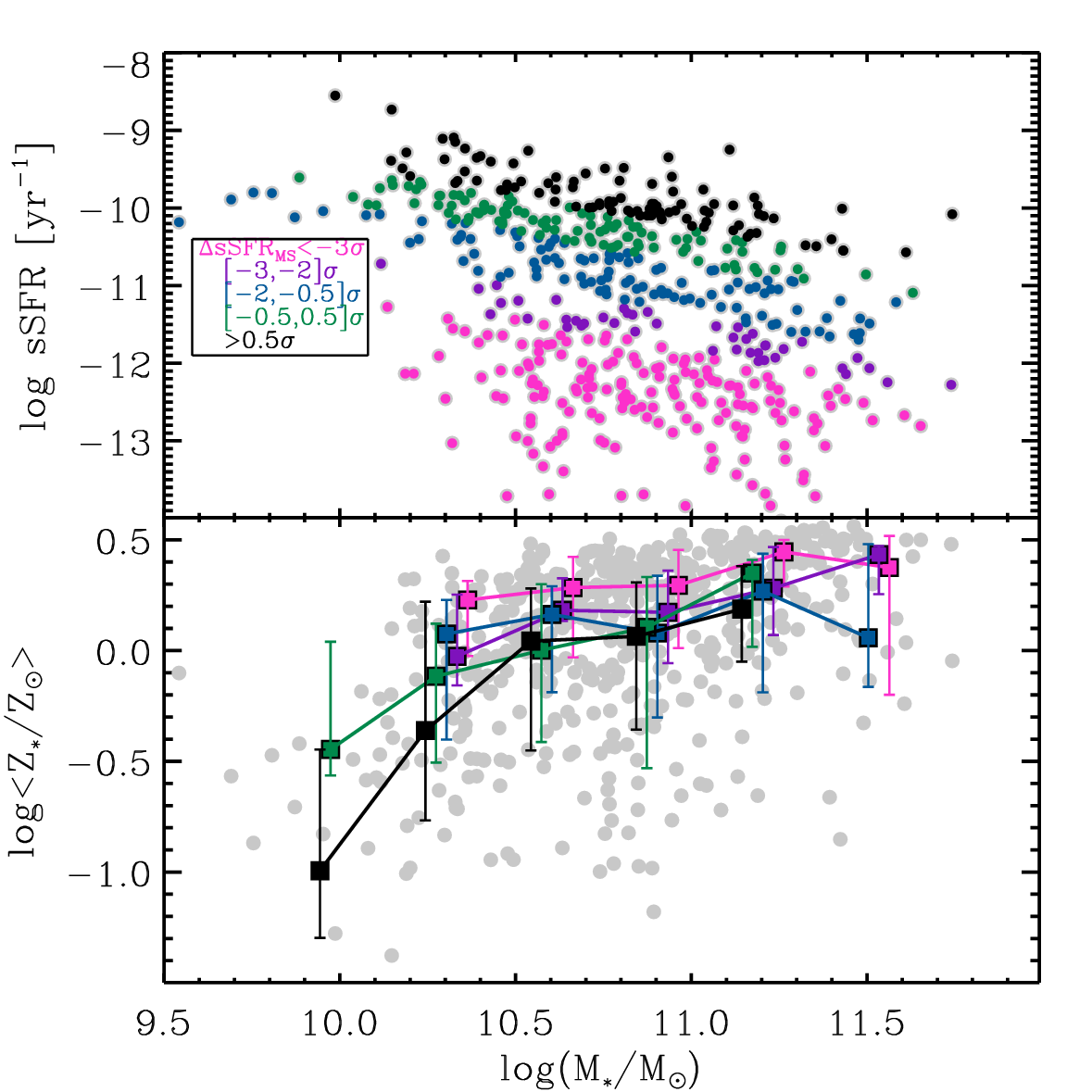}
    \includegraphics[width=0.4\linewidth]{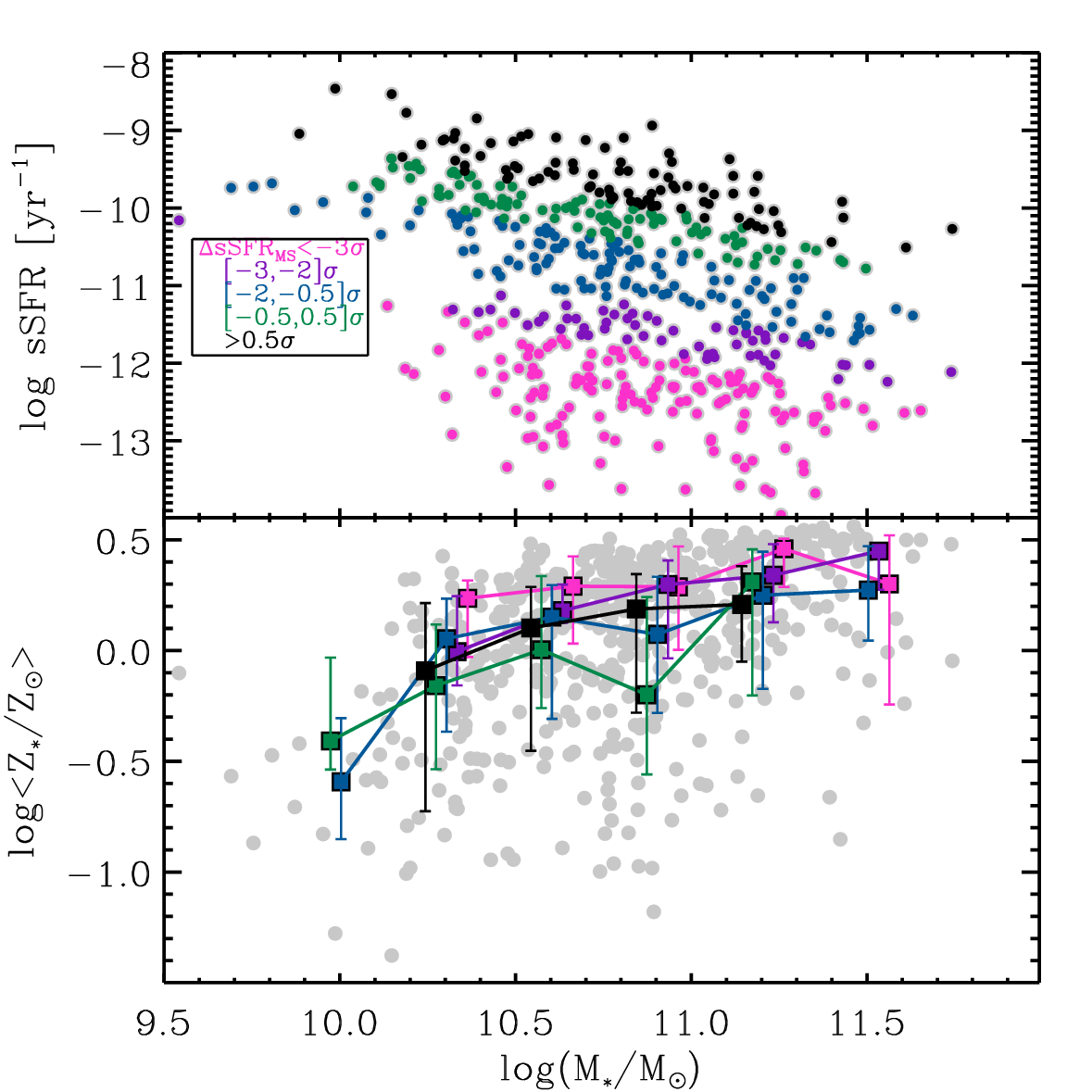}
    \caption{Comparison of the relation between stellar metallicity and SSFR at fixed mass as obtained with different SFR estimates. The upper panels display the SSFR versus stellar mass and the binning into distance from the Main Sequence; the bottom panels display the stellar metallicity-mass relations for bins of distance from the MS, as in Fig.~\ref{fig:scaling_sfr_legac}. These plots are based on SFR estimates from spectrum+photometry {\tt Prospector} fit \citep{Nersesian25}, instanteneous (left plot) or averaged over the previous 100 Myr (right plot).}
    \label{fig:compare_ssfr}
\end{figure}

The secondary dependence of stellar metallicity on SSFR at fixed mass (Fig.~\ref{fig:scaling_sfr_legac}) is also confirmed with any of the SFR estimates considered here, although with different significance. Figure~\ref{fig:compare_ssfr} compares the relations obtained with SFR estimates from spectroscopy+photometry {\tt Prospector} fits, instantaneous (left) or averaged over the past 100~Myr (right). Results based on {\tt Prospector} photometry-only fits are similar to the left panels. We consistently find that: i) galaxies with low stellar metallicity ($\log\left<Z_\ast/Z_\odot\right> <-0.5$) have the highest SSFR; ii) galaxies with low SSFR (i.e. below the MS) have all supersolar metallicity and a shallow relation with mass; iii) above $10^{11}\,\mathrm{M_\odot}$ we see no systematic dependence on SSFR; iv) below $10^{11}\,\mathrm{M_\odot}$ galaxies with higher SSFR are spread over a larger range in stellar metallicity (with increased downward scatter) than galaxies with lower SSFR. However, the choice of SFR estimate affects the strength of a continuous relation between $Z_\ast$ and SSFR at fixed stellar mass, and in particular the $Z_\ast-M_\ast$ relation which shifts upward for the highest $SSFR^{100}_\mathrm{Prosp,s}$ bin.

\begin{figure}
    \centering
    \includegraphics[width=0.4\linewidth]{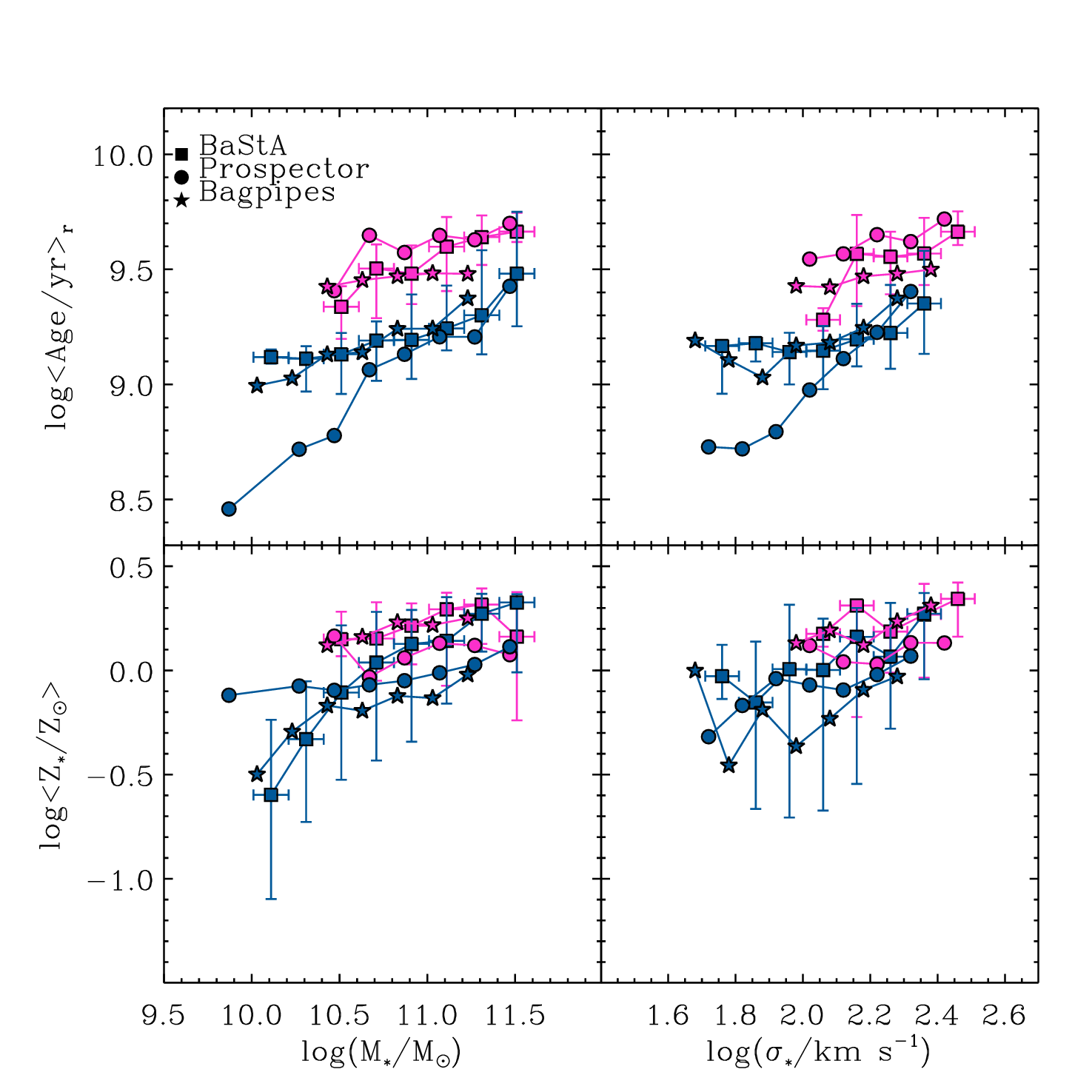}
    \caption{Median trends of light-weighted age (upper panels) and light-weighted stellar metallicity (lower panels) as a function of stellar mass (left) or velocity dispersion (right). The trends obtained with the {\tt BaStA} estimates (this work, squares) are compared with those obtained with parameter estimates from {\tt Prospector} \citep[][circles]{Nersesian25} and {\tt Bagpipes} \citep[][stars]{kaushal24}. Galaxies are distinguished into quiescent (magenta) and star-forming (blue) with our default $SSFR_\mathrm{UVIR}$ threshold. For clarity, the errobars show the $16^{th}-84^{th}$ range of the distributions only for {\tt BaStA} median trends. In this plot, metallicities have been scaled to a common solar scale of $Z_\odot=0.02$. Comparison is done for \golden~sample. }
    \label{fig:compare_relations_Q_SF}
\end{figure}

To test the impact of the spectral inference of physical parameters, we compare the median age and metallicity scaling relations for quiescent and star-forming galaxies derived in this work with {\tt BaStA} with those we would obtain, for the same sample, using estimates from {\tt Prospector} \citep{Nersesian25} and from {\tt Bagpipes} \citep{kaushal24} (see also Appendix E of Paper I). {\tt Prospector} estimates from \cite{Nersesian25} adopts Simple Stellar Populations constructed from FSPS with MILES stellar library and MIST isochrones, non-parametric SFHs and constant metallicity along the SFH (see their Table 1), while {\tt Bagpipes} estimates from \cite{kaushal24} adopts the 2016 version of \cite{bc03} models with MILES stellar library, double power-law SFH and constant metallicity (see their Table 1). 
In Fig.~\ref{fig:compare_relations_Q_SF} we show that all three codes agree in finding two distinct age sequences for quiescent and star-forming galaxies, with relations generally consistent  within 1-2$\sigma$~(where $\sigma$~is the error on the median from {\tt BaStA})  among the three codes. We note though that {\tt Prospector} assigns younger ages to low-mass star-forming galaxies with respect to {\tt BaStA} and {\tt Bagpipes}. The mass--metallicity relations also agree within 1$\sigma$: quiescent galaxies are more metal-rich than star-forming and follow a rather shallow relation with mass. Some differences are however noticeable: the difference between quiescent and star-forming galaxies is larger for {\tt Bagpipes}; {\tt Prospector} finds a flatter relation for star-forming galaxies and lower metallicities for quiescent galaxies with respect to the other estimates. Note that this is likely largely ascribed to the upper boundary at $0.2\cdot Z_\odot$~assumed in the {\tt Prospector} run,  following the conservative choice of safe parameter range by \cite{Leja19}.  This comparison serves to give a quantification of the cumulative effect of different model choices and ingredients on the derived physical parameters and their scaling relations. We note that the {\tt Bagpipes} and {\tt Prospector} runs are optimized to retrieve SFHs and timescales, while our {\tt BaStA} run, focusing on selected absorption features, is optimized for combined age and metallicity estimates. We thus consider the latter as reference metallicity estimates. 

\section{Median trends of age and stellar metallicity as a function of stellar mass and velocity dispersion}\label{appendix:tables}
Here we provide tables with the median trends shown in Fig.~\ref{fig:scaling_Q_SF_legac} and~\ref{fig:scaling_Q_SF_legac_sigma}. Table~\ref{Tab:median_relations_Q} reports the median and percentiles of the distribution in light-weighted age and stellar metallicity as a function of stellar mass for quiescent and star-forming galaxies in the \silver~and \golden~samples (shown in Fig.~\ref{fig:scaling_Q_SF_legac}. Table~\ref{Tab:median_relations_sigma} reports the median and percentiles of the distribution in light-weighted age and stellar metallicity as a function of velocity dispersion for quiescent and star-forming galaxies in the \silver~and \golden~samples (shown in Fig.~\ref{fig:scaling_Q_SF_legac_sigma}

\begin{table*}
\caption{Median trends of light-weighted age and stellar metallicity as a function of stellar mass as shown in Fig.\ref{fig:scaling_Q_SF_legac}.}\label{Tab:median_relations_Q}
\begin{center}
\begin{tabular}{|c|ccc|c|ccc|c|}
\hline \hline
\multicolumn{9}{|c|}{QUIESCENT galaxies} \\
\hline\hline
\multicolumn{9}{|c|}{\silver~sample, weighted by ${\tt Tcor}\times {\tt w\_spec\_silver}$} \\
\hline 
$\log(M_\ast/M_\odot$) & \multicolumn{3}{c|}{$\log<Age/yr>_r$} & $<2\times\sigma_{\log Age}>$  & \multicolumn{3}{c|}{$\log<Z_\ast/Z_\odot>_r$} &$<2\times\sigma_{\log Z_\ast}>$ \\
 & p50 & p16 & p84 &  & p50  & p16 & p84 & \\
\hline
\noalign{\vspace{3pt}}
 10.31 & $9.36^{0.06}_{0.08}$ &  9.21 &  9.56 & 0.35 &   $0.07^{0.03}_{0.09}$ &  $-0.01$ &  0.30 & 0.51 \\
 10.51 & $9.38^{0.05}_{0.04}$ &  9.15 &  9.55 & 0.35 & $0.24^{0.06}_{0.03}$ &  $-0.05$ &  0.37 & 0.44 \\
 10.71 & $9.47^{0.03}_{0.02}$ &  9.31 &  9.60 & 0.38 &  $0.27^{0.04}_{0.03}$ &  $ 0.07$ &  0.44 & 0.39 \\
 10.91 & $9.49^{0.04}_{0.03}$ &  9.30 &  9.64 & 0.36 & $0.30^{0.05}_{0.02}$ &  $ 0.04$ &  0.41 & 0.35 \\
 11.11 & $9.59^{0.04}_{0.02}$ &  9.39 &  9.72 & 0.32 &   $0.32^{0.06}_{0.03}$ &  $-0.02$ &  0.49 & 0.31 \\
 11.31 & $9.64^{0.02}_{0.02}$ &  9.52 &  9.75 & 0.31 &   $0.39^{0.04}_{0.02}$ &  $ 0.20$ &  0.51 & 0.29 \\
 11.51 & $9.69^{0.03}_{0.02}$ &  9.61 &  9.75 & 0.26 &   $0.21^{0.15}_{0.09}$ &  $-0.25$ &  0.49 & 0.23 \\
\noalign{\vspace{3pt}}
\hline
\multicolumn{9}{|c|}{\golden~sample} \\
\hline
$\log(M_\ast/M_\odot$) & \multicolumn{3}{c|}{$\log<Age/yr>_r$} & $<2\times\sigma_{\log Age}>$  & \multicolumn{3}{c|}{$\log<Z_\ast/Z_\odot>_r$} &$<2\times\sigma_{\log Z_\ast}>$ \\
 & p50 & p16 & p84 &  & p50 & p16 & p84 & \\
\hline
\noalign{\vspace{3pt}}
10.51 & $9.34^{0.06}_{0.04}$  & 9.20 & 9.43 & 0.32         & $0.27^{+0.03}_{-0.05}$ & 0.19 & 0.40 & 0.38 \\
10.71 & $9.50^{0.06}_{0.04}$  & 9.29 & 9.62 & 0.31         & $0.28^{+0.05}_{-0.05}$ & 0.11 & 0.45 & 0.38 \\
10.91 & $9.48^{0.03}_{0.03}$  & 9.36 & 9.60 & 0.36         & $0.30^{+0.05}_{-0.03}$ & 0.12 & 0.44 & 0.33 \\
11.11 & $9.60^{0.04}_{0.03}$  & 9.41 & 9.73 & 0.32         & $0.41^{+0.07}_{-0.02}$ & 0.04 & 0.49 & 0.27 \\
11.31 & $9.64^{0.03}_{0.02}$  & 9.52 & 9.73 & 0.31         & $0.43^{+0.04}_{-0.02}$ & 0.25 & 0.51 & 0.26 \\
11.51 & $9.66^{0.01}_{0.03}$  & 9.62 & 9.75 & 0.28         & $0.28^{+0.13}_{-0.07}$ & $-0.12$ & 0.49 & 0.24 \\
\noalign{\vspace{3pt}}
\hline \hline
\multicolumn{9}{|c|}{STAR-FORMING galaxies} \\
\hline \hline
\multicolumn{9}{|c|}{\silver~sample, weighted by ${\tt Tcor}\times {\tt w\_spec\_silver}$} \\
\hline
$\log(M_\ast/M_\odot$) & \multicolumn{3}{c|}{$\log<Age/yr>_r$} & $<2\times\sigma_{\log Age}>$  & \multicolumn{3}{c|}{$\log<Z_\ast/Z_\odot>_r$} &$<2\times\sigma_{\log Z_\ast}>$ \\
 & p50 & p16 & p84 &  & p50  & p16 & p84 & \\
\hline
\noalign{\vspace{3pt}}
        10.11 & $9.08^{+0.06}_{-0.03}$ & 8.88 & 9.19 & 0.29          & $-0.42^{+0.14}_{-0.05}$ & $-0.92$ & $-0.24$ & 0.93 \\
        10.31 & $9.14^{+0.04}_{-0.03}$ & 8.91 & 9.28 & 0.31          & $-0.05^{+0.08}_{-0.06}$ & $-0.46$ & 0.27 & 0.88 \\
        10.51 & $9.14^{+0.03}_{-0.02}$ & 9.00 & 9.25 & 0.32          & $0.03^{+0.07}_{-0.04}$ & $-0.34$ & 0.27 & 0.72 \\
        10.71 & $9.19^{+0.03}_{-0.01}$ & 9.01 & 9.28 & 0.32          & $0.13^{+0.09}_{-0.03}$ & $-0.43$ & 0.35 & 0.59 \\
        10.91 & $9.19^{+0.02}_{-0.04}$ & 9.05 & 9.47 & 0.31          & $0.02^{+0.10}_{-0.06}$ & $-0.62$ & 0.38 & 0.56 \\
        11.11 & $9.26^{+0.02}_{-0.03}$ & 9.15 & 9.41 & 0.32          & $0.23^{+0.08}_{-0.04}$ & $-0.21$ & 0.46 & 0.50 \\
        11.31 & $9.25^{+0.06}_{-0.08}$ & 9.04 & 9.53 & 0.29          & $0.36^{+0.09}_{-0.02}$ & 0.03 & 0.45 & 0.38 \\
        11.51 & $9.28^{+0.08}_{-0.21}$ & 9.08 & 9.75 & 0.29          & $0.30^{+0.44}_{-0.08}$ & $-0.69$ & 0.48 & 0.31 \\
\noalign{\vspace{3pt}}
\hline
\multicolumn{9}{|c|}{\golden~sample} \\
\hline
$\log(M_\ast/M_\odot$) & \multicolumn{3}{c|}{$\log<Age/yr>_r$} & $<2\times\sigma_{\log Age}>$  & \multicolumn{3}{c|}{$\log<Z_\ast/Z_\odot>_r$} &$<2\times\sigma_{\log Z_\ast}>$ \\
 & p50 & p16 & p84 &  & p50 & p16 & p84 & \\
\hline
\noalign{\vspace{3pt}}
        10.11 & $9.12^{+0.01}_{-0.01}$ & 9.09 & 9.15 & 0.26          & $-0.48^{+0.22}_{-0.16}$ & $-0.98$ & $-0.12$ & 0.70 \\
        10.31 & $9.11^{+0.05}_{-0.02}$ & 8.97 & 9.17 & 0.31          & $-0.21^{+0.15}_{-0.10}$ & $-0.61$ & 0.07 & 0.89 \\
        10.51 & $9.13^{+0.04}_{-0.02}$ & 8.96 & 9.22 & 0.30          & $0.01^{+0.10}_{-0.08}$ & $-0.41$ & 0.33 & 0.68 \\
        10.71 & $9.19^{+0.04}_{-0.02}$ & 9.02 & 9.27 & 0.30          & $0.16^{+0.10}_{-0.05}$ & $-0.31$ & 0.40 & 0.55 \\
        10.91 & $9.19^{+0.04}_{-0.04}$ & 9.02 & 9.39 & 0.31          & $0.25^{+0.10}_{-0.03}$ & $-0.22$ & 0.41 & 0.49 \\
        11.11 & $9.24^{+0.02}_{-0.04}$ & 9.15 & 9.43 & 0.32          & $0.26^{+0.07}_{-0.05}$ & $-0.04$ & 0.47 & 0.38 \\
        11.31 & $9.30^{+0.05}_{-0.09}$ & 9.13 & 9.58 & 0.29          & $0.39^{+0.06}_{-0.03}$ & 0.21 & 0.49 & 0.38 \\
        11.51 & $9.48^{+0.11}_{-0.13}$ & 9.25 & 9.75 & 0.29          & $0.44^{+0.16}_{-0.02}$ & 0.11 & 0.48 & 0.28 \\
\noalign{\vspace{3pt}}
\hline
\end{tabular}
\tablefoot{Median (p50) and percentiles (p16, p84) of the light-weighted age and stellar metallicity in bins of stellar mass for quiescent and star-forming galaxies belonging to the \silver~sample (weighted for volume and spectroscopic completeness) and to the \golden~sample. The mean uncertainties, expressed as twice the interpercentile range of the parameter PDF, are also provided ($<2\times\sigma_{\log Age}>$, $<2\times\sigma_{\log Z_\ast}>$ ).}
\end{center}
\label{Tab:mean_relations}
\end{table*}

\begin{table*}
\caption{Median trends of light-weighted age and stellar metallicity as a function of velocity dispersion as shown in Fig.\ref{fig:scaling_Q_SF_legac_sigma}.}\label{Tab:median_relations_sigma}
\begin{center}
\begin{tabular}{|c|ccc|c|ccc|c|}
\hline \hline
\multicolumn{9}{|c|}{QUIESCENT galaxies} \\
\hline\hline
\multicolumn{9}{|c|}{\silver~sample, weighted by ${\tt Tcor}\times {\tt w\_spec\_silver}$} \\
\hline 
$\log \sigma_\ast [\rm km/s$] & \multicolumn{3}{c|}{$\log<Age/yr>_r$} & $<2\times\sigma_{\log Age}>$  & \multicolumn{3}{c|}{$\log<Z_\ast/Z_\odot>_r$} &$<2\times\sigma_{\log Z_\ast}>$ \\
 & p50 & p16 & p84 &  & p50  & p16 & p84 & \\
\hline
\noalign{\vspace{3pt}}
        2.06 & $9.31^{+0.03}_{-0.03}$ & 9.23 & 9.39 & 0.35         & $-0.01^{+0.18}_{-0.10}$ & $-0.54$ & 0.30 & 0.51 \\
        2.16 & $9.38^{+0.06}_{-0.06}$ & 9.10 & 9.65 & 0.38         & $0.27^{+0.04}_{-0.03}$ & 0.07 & 0.42 & 0.44 \\
        2.26 & $9.53^{+0.03}_{-0.01}$ & 9.35 & 9.61 & 0.35         & $0.18^{+0.03}_{-0.02}$ & $-0.04$ & 0.35 & 0.43 \\
        2.36 & $9.57^{+0.03}_{-0.02}$ & 9.38 & 9.70 & 0.32         & $0.39^{+0.05}_{-0.01}$ & 0.08 & 0.48 & 0.28 \\
        2.46 & $9.66^{+0.03}_{-0.02}$ & 9.54 & 9.72 & 0.31         & $0.33^{+0.05}_{-0.04}$ & 0.13 & 0.50 & 0.30 \\
\noalign{\vspace{3pt}}
\hline
\multicolumn{9}{|c|}{\golden~sample} \\
\hline
$\log \sigma_\ast [\rm km/s$] & \multicolumn{3}{c|}{$\log<Age/yr>_r$} & $<2\times\sigma_{\log Age}>$  & \multicolumn{3}{c|}{$\log<Z_\ast/Z_\odot>_r$} &$<2\times\sigma_{\log Z_\ast}>$ \\
 & p50 & p16 & p84 &  & p50 & p16 & p84 & \\
\hline
\noalign{\vspace{3pt}}
        2.06 & $9.28^{+0.02}_{-0.02}$ & 9.23 & 9.33 & 0.31         & $0.29^{+0.03}_{-0.07}$ & 0.23 & 0.44 & 0.37 \\
        2.16 & $9.54^{+0.07}_{-0.08}$ & 9.36 & 9.74 & 0.32         & $0.44^{+0.06}_{-0.03}$ & 0.30 & 0.51 & 0.24 \\
        2.26 & $9.53^{+0.03}_{-0.02}$ & 9.35 & 9.65 & 0.34         & $0.26^{+0.04}_{-0.03}$ & 0.04 & 0.43 & 0.36 \\
        2.36 & $9.61^{+0.02}_{-0.02}$ & 9.49 & 9.72 & 0.31         & $0.41^{+0.05}_{-0.01}$ & 0.08 & 0.49 & 0.27 \\
        2.46 & $9.66^{+0.03}_{-0.02}$ & 9.54 & 9.74 & 0.30         & $0.45^{+0.05}_{-0.02}$ & 0.28 & 0.53 & 0.23 \\
\noalign{\vspace{3pt}}
\hline \hline
\multicolumn{9}{|c|}{STAR-FORMING galaxies} \\
\hline \hline
\multicolumn{9}{|c|}{\silver~sample, weighted by ${\tt Tcor}\times {\tt w\_spec\_silver}$} \\
\hline
$\log \sigma_\ast [\rm km/s$] & \multicolumn{3}{c|}{$\log<Age/yr>_r$} & $<2\times\sigma_{\log Age}>$  & \multicolumn{3}{c|}{$\log<Z_\ast/Z_\odot>_r$} &$<2\times\sigma_{\log Z_\ast}>$ \\
 & p50 & p16 & p84 &  & p50  & p16 & p84 & \\
\hline
\noalign{\vspace{3pt}}
        1.76 & $9.07^{+0.00}_{-0.05}$ & 9.06 & 9.19 & 0.25         & $-0.41^{+0.06}_{-0.26}$ & $-0.54$ & 0.17 & 0.86 \\
        1.86 & $9.10^{+0.05}_{-0.02}$ & 8.91 & 9.20 & 0.28         & $-0.46^{+0.19}_{-0.18}$ & $-1.23$ & 0.25 & 0.78 \\
        1.96 & $9.10^{+0.02}_{-0.04}$ & 9.00 & 9.30 & 0.31         & $-0.07^{+0.11}_{-0.06}$ & $-0.62$ & 0.25 & 0.73 \\
        2.06 & $9.15^{+0.03}_{-0.02}$ & 8.98 & 9.26 & 0.31         & $-0.11^{+0.06}_{-0.06}$ & $-0.49$ & 0.31 & 0.79 \\
        2.16 & $9.13^{+0.02}_{-0.03}$ & 9.02 & 9.34 & 0.32         & $-0.05^{+0.06}_{-0.05}$ & $-0.44$ & 0.33 & 0.73 \\
        2.26 & $9.22^{+0.03}_{-0.02}$ & 9.04 & 9.37 & 0.31         & $-0.01^{+0.08}_{-0.05}$ & $-0.55$ & 0.34 & 0.62 \\
        2.36 & $9.33^{+0.09}_{-0.08}$ & 9.01 & 9.63 & 0.34         & $0.25^{+0.09}_{-0.05}$ & $-0.08$ & 0.44 & 0.37 \\
\noalign{\vspace{3pt}}
\hline
\multicolumn{9}{|c|}{\golden~sample} \\
\hline
$\log \sigma_\ast [\rm km/s$] & \multicolumn{3}{c|}{$\log<Age/yr>_r$} & $<2\times\sigma_{\log Age}>$  & \multicolumn{3}{c|}{$\log<Z_\ast/Z_\odot>_r$} &$<2\times\sigma_{\log Z_\ast}>$ \\
 & p50 & p16 & p84 &  & p50 & p16 & p84 & \\
\hline
\noalign{\vspace{3pt}}
        1.86 & $9.18^{+0.03}_{-0.01}$ & 9.10 & 9.21 & 0.29         & $-0.04^{+0.18}_{-0.10}$ & $-0.55$ & 0.26 & 0.79 \\
        1.96 & $9.14^{+0.04}_{-0.03}$ & 9.00 & 9.23 & 0.28         & $0.23^{+0.25}_{-0.06}$ & $-0.64$ & 0.44 & 0.60 \\
        2.06 & $9.14^{+0.04}_{-0.02}$ & 8.96 & 9.25 & 0.30         & $0.12^{+0.13}_{-0.05}$ & $-0.54$ & 0.36 & 0.62 \\
        2.16 & $9.20^{+0.02}_{-0.03}$ & 9.08 & 9.35 & 0.30         & $0.28^{+0.14}_{-0.03}$ & $-0.43$ & 0.42 & 0.49 \\
        2.26 & $9.22^{+0.03}_{-0.04}$ & 9.07 & 9.43 & 0.31         & $0.18^{+0.07}_{-0.05}$ & $-0.16$ & 0.44 & 0.44 \\
        2.36 & $9.37^{+0.09}_{-0.07}$ & 9.10 & 9.59 & 0.34         & $0.39^{+0.10}_{-0.03}$ & 0.07 & 0.48 & 0.34 \\
\noalign{\vspace{3pt}}
\hline
\end{tabular}
\tablefoot{Median (p50) and percentiles (p16, p84) of the light-weighted age and stellar metallicity in bins of stellar velocity dispersion for quiescent and star-forming galaxies belonging to the \silver~sample (weighted for volume and spectroscopic completeness) and to the \golden~sample. The mean uncertainties, expressed as twice the interpercentile range of the parameter PDF, are also provided ($<2\times\sigma_{\log Age}>$, $<2\times\sigma_{\log Z_\ast}>$ ).}
\end{center}
\end{table*}

\end{appendix}

\end{document}